\documentclass[8.5pt,twoside,twocolumn]{article}
\usepackage{times,mathptmx}
\usepackage{amssymb,amsfonts,amsmath}
\usepackage[square,sort,comma,numbers]{natbib}
\usepackage{graphicx} 
\usepackage{lastpage}
\usepackage{dcolumn}
\begin{document}

\thispagestyle{plain}
\renewcommand{\thefootnote}{\fnsymbol{footnote}}
\renewcommand\footnoterule{\vspace*{1pt}%
\hrule width 3.4in height 0.4pt \vspace*{5pt}} 
\setcounter{secnumdepth}{5}

\makeatletter 
\def\subsubsection{\@startsection{subsubsection}{3}{10pt}{-1.25ex plus -1ex minus -.1ex}{0ex plus 0ex}{\normalsize\bf}} 
\def\paragraph{\@startsection{paragraph}{4}{10pt}{-1.25ex plus -1ex minus -.1ex}{0ex plus 0ex}{\normalsize\textit}} 
\renewcommand\@biblabel[1]{#1}            
\renewcommand\@makefntext[1]%
{\noindent\makebox[0pt][r]{\@thefnmark\,}#1}
\makeatother 
\renewcommand{\figurename}{\small{Fig.}~}
\newcommand{\pre}{Phys. Rev. E}
\newcommand{\prb}{Phys. Rev. B}
\newcommand{\pra}{Phys. Rev. A}
\newcommand{\prl}{Phys. Rev. Lett.}
\newcommand{\rmp}{Rev. Mod. Phys.}


\twocolumn[
  \begin{@twocolumnfalse}
\noindent\LARGE{\textbf{Pre-yield non-affine fluctuations and a hidden critical point in strained crystals$^\dag$}}
\vspace{0.6cm}

\noindent\large{\textbf{Tamoghna Das,\textit{$^{a,b}$} Saswati Ganguly,\textit{$^{b}$} Surajit Sengupta\textit{$^{\ast}$}\textit{$^{c}$} and Madan Rao\textit{$^{d\ddag}$}}}\vspace{0.5cm}

\noindent{\textit{$^{a}$~Collective Interactions Unit, OIST Graduate University,1919-1 Tancha, Onna-son, Okinawa, Japan - 904-0495}}

\noindent{\textit{$^{b}$~Centre for Advanced Materials, Indian Association for the Cultivation of Science, Jadavpur, Kolkata 700032, India }}

\noindent{\textit{$^{c}$~TIFR Centre for Interdisciplinary Sciences, 21 Brundavan Colony, Narsingi, Hyderabad 500075, India.}\,email:\,surajit@tifrh.res.in}

\noindent{\textit{$^{d}$~Raman Research Institute, C.V. Raman Avenue, Bangalore 560080, India }}

%
\vspace{0.6cm}

\noindent \normalsize {A crystalline solid exhibits thermally induced localised {\em non-affine} droplets in the absence of external stress. Here we show that upon an imposed shear, the size of these droplets grow until they percolate at a critical strain, well {\em below} the value at which the solid begins to yield. This critical point does not manifest in bulk thermodynamic or mechanical properties, but is {\em hidden} and reveals itself in the onset of inhomogeneities in elastic moduli, marked changes in the appearance and local properties of non-affine droplets and a sudden enhancement in defect pair concentration. Slow relaxation of stress and an-elasticity appear as observable dynamical consequences of this hidden criticality. Our results may be directly verified in colloidal crystals with video microscopy techniques but are expected to have more general validity. 
}
\vspace{0.5cm}
 \end{@twocolumnfalse}
]

\footnotetext{\dag~Electronic Supplementary Information (ESI) available: [attached file: Supplementary text]}



\footnotetext{\ddag~Also at National Centre for Biological Sciences (TIFR), Bellary Road, Bangalore 560065, India
}


Mechanical properties of solids~\cite{dieter,haasen}, especially mechanisms of yielding in response to external stress, continue to engage the attention of materials scientists and engineers~\cite{fcc-plast}. Pre-yield phenomena such as anelasticity, occurring at non-zero temperatures and below the yield stress, in both crystals~\cite{anelasticity} and metallic glasses~\cite{ana-glass1,ana-glass2}  is somewhat less understood, however. Within this anelastic regime,  a crystalline solid undergoes recoverable strain but with a long relaxation time and is accompanied by conspicuous production and reorganisation of lattice defects~\cite{Hirth, ane2} marked by the onset of significant departure from instantaneous and linear, ``Hooke's law'', elasticity. Mechanical response becomes {\em heterogeneous}; the regions with defects behaving differently from the rest of the solid. In amorphous solids too, anelasticity is marked by the appearance of similar heterogeneous mechanical response confined to localised clusters of particles undergoing large {\em non-affine} deformation or non affine droplets~\cite{Argon,spaepen,FL}. 

In this paper, we focus on defect nucleation and associated pre-yield phenomena in two (2D) and three-dimensional (3D), initially homogeneous, crystals of interacting particles.    
We provide strong evidence associating the proliferation of crystalline defects during deformation at non-zero temperatures, with the critical behaviour of thermally generated non-affine droplets. These non-affine displacements exist in crystalline solids even in the absence of an applied strain and have a purely thermal origin \cite{saswati}. Upon shearing the crystalline solid, these clusters grow and eventually percolate at a hidden mechanical critical point; concomitantly, defects suddenly proliferate. Significantly, this happens well before the yield point of the crystalline solid. 
Apart from its intrinsic novelty, this viewpoint sheds insight onto a broader set of issues. First, we find that dislocation nucleation is stochastic and seeded from regions with large non-affine displacements.  Second, we have identified the localised non-affine regions with droplet excitations from nearby metastable liquid/glass~\cite{pre}. Thus,  our finding that the percolation of non-affine droplets coincides with the disappearance of the metastable liquid-glass spinodal, brings out the significance of metastable configurations in understanding mechanical properties of solids. Last, this perspective might provide a language bridging ideas concerning the mechanical response of crystals and amorphous solids.
\begin{figure*}[ht]
\begin{center}
\includegraphics[width= 0.7\textwidth]{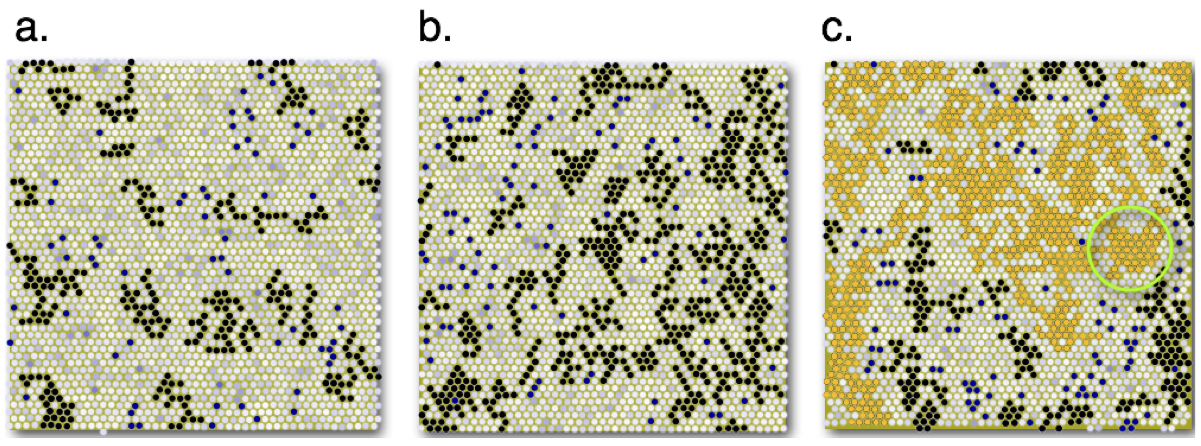}
\end{center}
\caption{ (color-online) {\bf a}-{\bf c}\, Configurations from a part of our constant $NVT$, MD simulation cell of $22500$ 2D-LJ particles at $T = 0.4$, $\rho = 0.99$ with $\epsilon = 0$ ({\bf a}). Shading (light to dark) tracks local $\chi$. Non-affine particles ($\chi > \chi_{cut}$), organised into droplets, are black. Strain increases non-affine particles; $\epsilon = 0.04$ ({\bf b}) and droplets percolate at $\epsilon = 0.052$ ({\bf c}). Percolating droplet in yellow (dark grey) with a compact node circled green (light grey). }
\label{pchi}

\end{figure*}

The existence of a unique reference in crystals allows us to identify localized non-affine regions at non-zero temperatures, even at zero applied strain. In Ref.\cite{pre} we studied the statistics of shape and size of non-affine clusters and their local thermodynamic properties using molecular dynamics (MD) simulations of a two-dimensional Lennard-Jones (2D-LJ) solid \cite{ums,LJ}. We summarise the main results below for completeness.

For every configuration, particles $i$ undergoing large non-affine displacements are identified using the parameter $\chi_i$ viz. the least square error incurred in trying to {\em fit} an instantaneous local volume, $\Omega$, to an affine distortion of the same volume in the, reference, undistorted lattice~\cite{FL}. A cutoff criterion $\chi_i > \chi_{cut}$ then eliminates trivial {\em harmonic} distortions of the lattice (see {\bf Methods} and supplementary information SI). These particles are observed to cluster together into droplets.
%
The droplets are characterized by a distribution of the local density $\rho_c$ and excess pressures, $\Delta p_c \equiv p_{c}-p$. Here $p$ is the mean pressure of the surrounding solid of density $\rho$ and $p_c$ is computed by averaging the virial over the $n_c$ particles of the cluster. For any temperature $T$ the mean scaled excess pressure ${\bar p} = \Delta p_c/T$ vs $\rho_c$ curve is non-monotonic and resembles a van der Waals loop seen in usual liquid-gas transitions. The two stable branches with $\partial {\bar p}/\partial \rho_c > 0$ are connected by an {\em ``unstable''} branch where $\partial {\bar p}/\partial \rho_c < 0$.  In the unstrained solid the stable branches were associated with inflated, compact clusters or deflated, string-like clusters depending on whether ${\bar p}$  is negative or positive respectively\,\cite{LSF,mags}; the unstable branch contains only ramified ``branched-polymer'' clusters \cite{brpl}. By studying the local equation of state and density-correlations, one may associate compact and string-like clusters with droplet fluctuations from nearby liquid and glass -like metastable free-energy minima respectively \cite{pre}. As $T$ is increased, the van der Waals loop as well as the distinction between compact and string-like droplets vanish above a {\em metastable} critical point. This behaviour is quite general and exists in other model systems admitting non-crystalline phases, (see discussion in SI). 

Here, we extend this earlier analysis \cite{pre} to the {\em mechanical response} of LJ solids in two and three dimensions over a range of  $\rho$, and for a few $T$. In what follows, the bulk of our quantitatively detailed results, unless otherwise stated, are explicitly demonstrated for the 2D-LJ solid for which computations, as well as experimental verification using real-time video microscopy~\cite{zahn}, are relatively cheap. Key results for the 3D-LJ case~\cite{3dlj3pt} are also presented showing that our main conclusions carry over to higher dimensions. 


\section*{Results}
\subsection*{Percolation of clusters}
As the crystal is subject to a {\em quasi-static} (pure) shear strain $\epsilon = \epsilon_{xx} - \epsilon_{yy}$ (Fig.\ref{pchi}a-c) (see {\bf Methods}), localized non-affine deformations grow and increase in number. 
At a critical value of the strain $\epsilon^* ({\rm e.g.}\,\approx 0.05 \,{\rm for}\, \rho = 0.99, T = 0.4)$, regions containing non-affine particles begin to percolate.
Further increase of the strain beyond $\epsilon^*$, eventually leads to yielding of the solid (at $\epsilon \approx 0.1\,{\rm for\, same}\,\rho, T)$. 

\begin{figure}[h!]
\begin{center}
\includegraphics[width=0.48\textwidth]{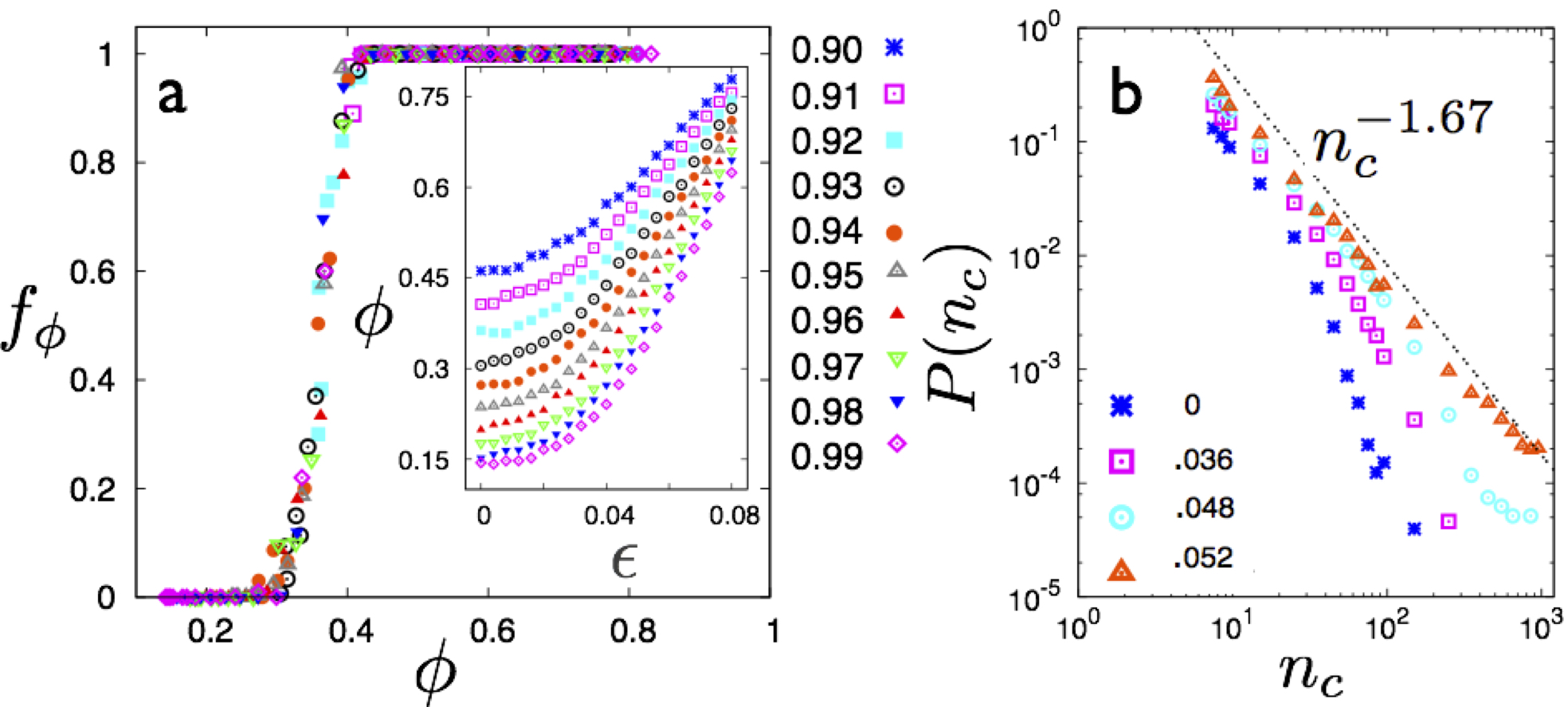}
\end{center}
 \caption{{\bf a} Scaling plot of $f_\phi$ vs. $\phi$ (see text) for various $\rho$ and $\epsilon$ showing data collapse for the 2D-LJ solid.  The percolation threshold $\phi^* \lesssim 0.4$. {\it Inset} Plots of $\phi$ vs $\epsilon$ for various $\rho$. {\bf b}  Probability $P(n_c)$   for $\rho = 0.99$ and $\epsilon = 0,.036,.048$ and $.052$ showing crossover from exponential to power-law behaviour at percolation $\epsilon^* \approx .052$; fitted exponent $\tau = -1.67$ of $P(n_c)=n_c^{-\tau}$ shown by dotted straight line. 
 }
\label{perc}
\end{figure}

To ascertain the nature of this percolation transition, we plot  the ratio of the number of particles in the largest cluster to the total number of non-affine particles, $f_\phi$ in the 2D-LJ solid as a function of the fraction of non-affine particles $\phi(\rho, \epsilon)$ in Fig.\ref{perc}a. The collapse of $f_\phi$ onto a single curve for all $\rho$ and $\epsilon$, is a strong indication  of a true critical percolation transition at a value of $\phi^* \simeq 0.4$, close to the known value of site percolation in 2D ($0.34$) \cite{percolation}. 
We plot the probability distribution $P(n_c)$ of obtaining a non-affine cluster of size $n_c$ for increasing values of $\epsilon$ in Fig.\ref{perc}b. As one approaches the percolation transition, $P(n_c) \sim n_c^{-\tau}$ with an exponent $\tau \approx 1.67$.

We now ask whether there is a signature of this geometrical critical point in the thermodynamics and whether these non-affine clusters admit a thermodynamic interpretation. The answer to this is surprisingly subtle.

\begin{figure}[h!]
\begin{center}
\includegraphics[width=0.47\textwidth]{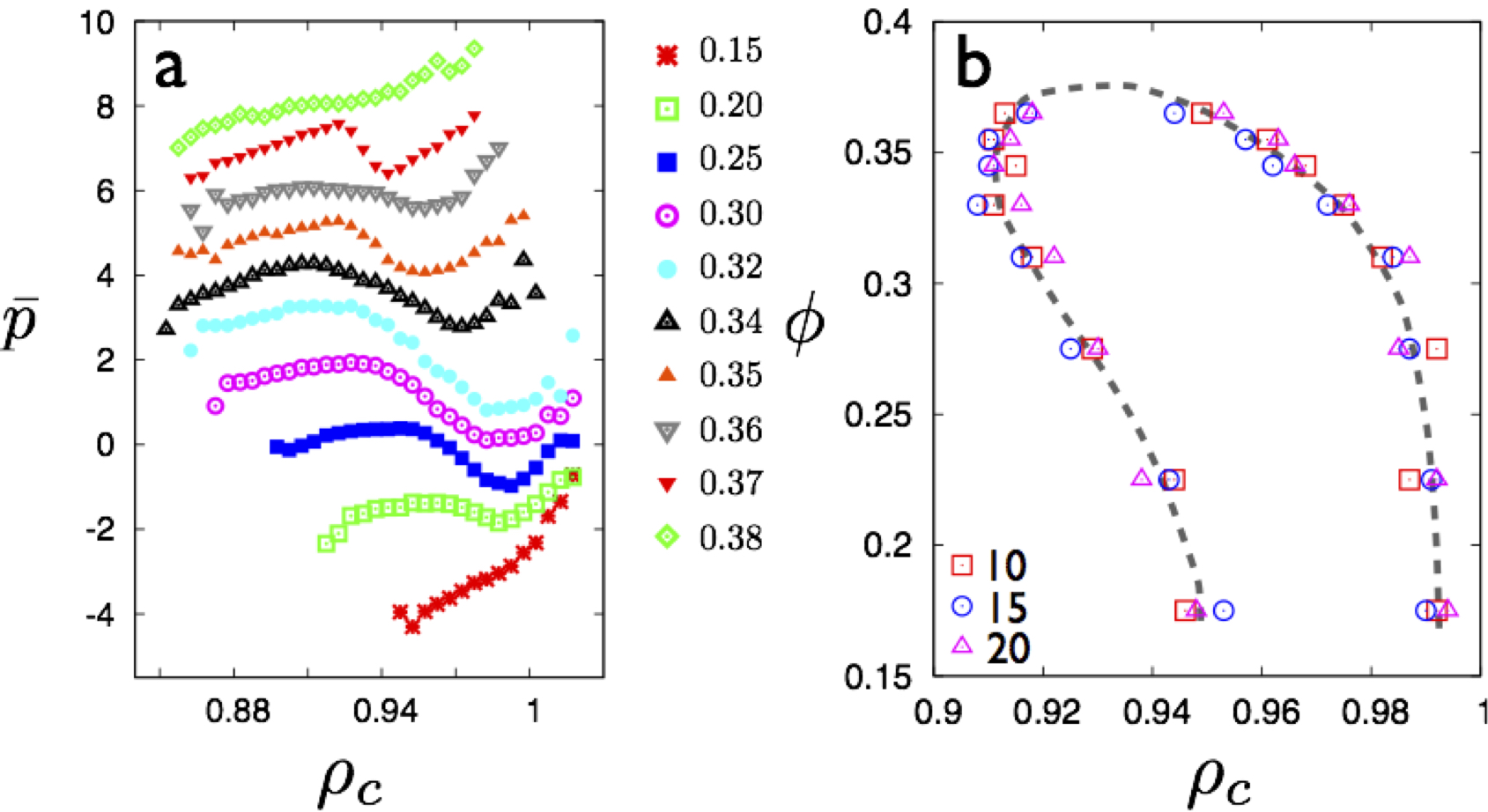}
\end{center}
 \caption{ {\bf a} Plot of  $\bar{p}$ vs. $\rho_c$ for various $\phi$ (key on right) and $n_c=15$ showing van der Waals loops. Loops for different $\phi$ have been shifted in $\bar{p}$-axis for clarity. Note that the loop disappears for $\phi > \phi^*$. {\bf b} Plot of the spinodal obtained for $n_c = 10, 15$ and $20$ in the $\phi- \rho_c$ plane clearly indicating a metastable critical point coinciding with $\phi^*$.
}
\label{perc1}
\end{figure}

\subsection*{Local thermodynamics} 
Bulk thermodynamic quantities such as equation of state etc. show no abrupt change across the transition.  
We, therefore, focus on measuring  local thermodynamic variables {\em within} the non-affine clusters and monitoring them across the percolation transition. We characterise droplets using the local quantities $\bar p$ and $\rho_c$ as in ~\cite{pre} (see also {\bf Methods}). The local temperatures, on the other hand, are close to the fixed temperature of the solid and are irrelevant to the subsequent analysis. Instead of temperature here strain, $\epsilon$, drives the percolation transition. 

We anticipate that systems with various $\epsilon$ and $\rho$ behave similarly for equal $\phi$. We test this as follows. From configurations with fraction of non-affine particles between $\phi$ and $\phi + \Delta \phi$, where $\Delta \phi$ is a suitable bin-size, we obtain the mean $\bar{p}$ and $\rho_c$ corresponding to droplets with fixed number of particles $10 < n_c <  100$~; collecting together data from all solids with $0.91 \leq \rho \leq 0.99$ and $\epsilon$ below the yield point. This data is plotted in Fig.\ref{perc1}a for droplets of size $n_c = 15$ for each value of $\phi$. Similar plots are obtained for droplets with other values of $n_c$ too. 

Over a large range of $\phi$ values, the plots are non-monotonic, showing prominent van der Waals loops with two stable branches. We show below that these corresponding to compact and string-like droplets. As $\phi$ increases, the  loops tend to vanish. Within the accuracy of our computations, the value of $\phi$ above which the loops vanish is the same as $\phi^*$, i.e. the non-affine fraction at which the clusters percolate.  To make this identification quantitative, we obtain the values of $\rho_c$ for all $\phi$ at the spinodal $\partial {\bar p}/\partial \rho_c = 0$ and plot them in Fig.\ref{perc1}b for $n_c = 10,15$ and $20$. When extrapolated, the spinodal lines intersect at a metastable, droplet critical point. This metastable critical point and the droplet shape transition survives for even higher values of $n_c$, though for very large $n_c$, the statistics becomes sparse. 
\begin{figure}[h!]
\centerline{\includegraphics[width=0.54\textwidth]{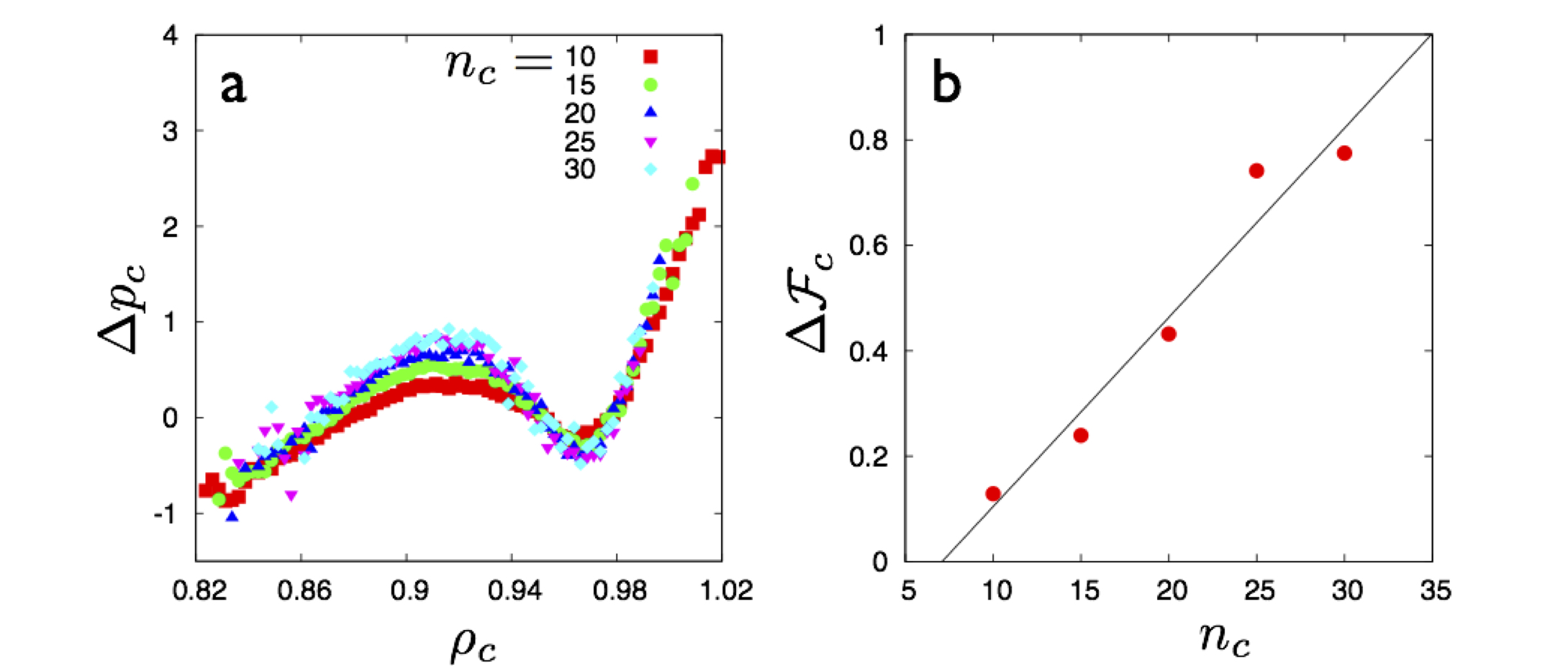}}
\caption{
{\bf a} Plot showing $\Delta p_c$ as a function of the local density $\rho_c$ for ``non-affine'' droplets in the strain free LJ solid at $T = 0.4$. The different colors show the van der Waals loop for various sizes of clusters $10< n_c < 30$. The data becomes sparse for much higher values of $n_c$, leading to worse statistics. The van der Waals loops however show stabilisation with larger $n_c$. We have made this quantitative by obtaining an effective barrier height $\Delta {\mathcal F}_c$ for this transition by integrating the pressure difference $\Delta p_c$ along $\rho_c$ for each of these curves. The result is plotted in ${\bf b}$ and shows that $\Delta {\mathcal F}_c$ increases with $n_c$ as expected of a first order transition. The solid line is a linear fit.}
\label{FSS}
\end{figure}

How does the extent of the van der Waals loop in $\bar p$ vs. $\rho_c$ depend on the size of the cluster $n_c$ ? In Fig.~\ref{FSS}a we show a comparison of the van der Waals loops for various values of $n_c$ in the $\epsilon \to 0$ limit.  An integration of $\bar p$ vs $\rho_c$ which gives a measure of the free energy ${\mathcal F}_c$ expended by the solid to produce these fluctuations has a convex up region, showing that an interpretation in terms of a metastable first order transition is not inconsistent. A metastable  first order transition implies, at least for compact droplets, that the surface free energy barriers should scale as $l_c^{d-1}$, where $l_c$ is a typical linear size associated with the clusters and $d (=2)$ is the dimensionality \cite{binder}. The actual form of the scaling may be more complicated because of the complex shape of the clusters. Nevertheless, in Fig.\ref{FSS}b we show a plot of the barrier height $\Delta {\mathcal F}_c$, obtained by integrating Fig.\ref{FSS}a, as a function of the size of the fluctuation. A monotonically increasing curve indicates a positive surface energy of the droplets at $\epsilon = 0$. At the metastable critical point we expect this surface energy to vanish, although large fluctuations and strong finite size effects makes an explicit evaluation of this quantity near $\phi = \phi^*$ computationally impractical. 

\subsection*{The shape transition}
We have shown that the van der Waals loop in ${\bar p}$ vs $\rho_c$ vanishes as $\phi \geq \phi^*$ where the non-affine droplets percolate. The van der Waals loop corresponds to a metastable transition between droplets which come in two stable shapes, either compact or string-like. We show now that this distinction between these shapes also vanishes for $\phi > \phi^*$.  Above this transition, string-like droplets connect the adjacent compact ``nodes'' in a single giant percolating network. This network has an overall, branched polymer shape so that isolated compact and string-like droplets cease to exist beyond $\phi  \geq \phi^*$ (see Fig.\ref{pchi}\,c). 
\begin{figure}[h]
\centerline{\includegraphics[width=0.4\textwidth]{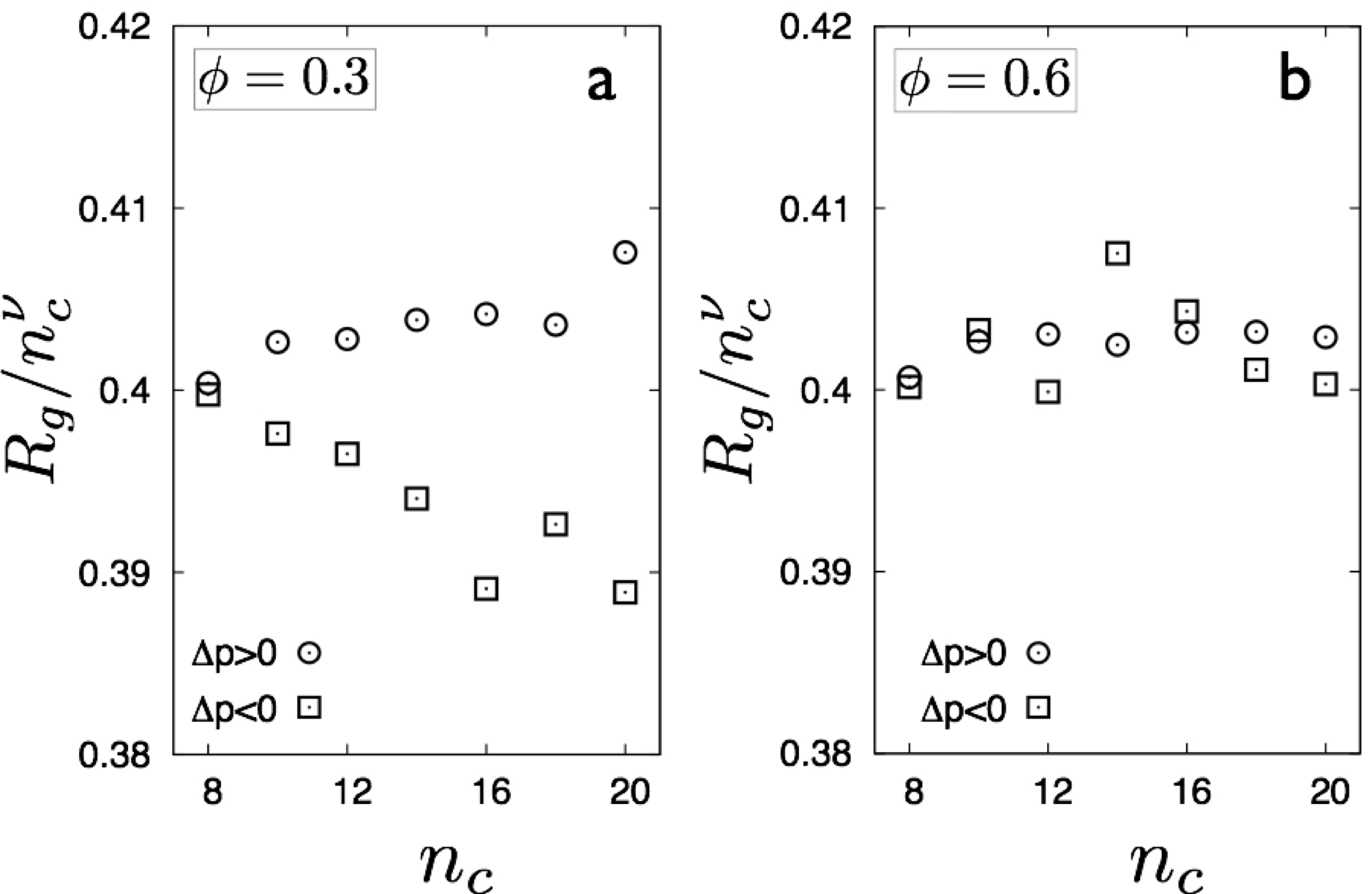}}
\caption{
The radius of gyration $R_g$ as a function of the size $n_c$ of the droplets where we have factored out the $\sim n_c^{\nu}$, with $\nu = 0.64$ (branched polymer) dependence. Below the percolation transition, {\bf a}, the droplets for the $\bar p > 0$ and $\bar p < 0$ branches behave differently assuming string-like ($\nu \sim 1$) and compact ($\nu \sim 1/2$) configurations respectively. Above percolation {\bf b}, the morphology of droplets changes drastically and all droplets become branched polymers.
}
\label{rgvsnc}
\end{figure}  

For the unstrained solid \cite{pre} the ${\bar p} < 0$ and ${\bar p} > 0$ branches of the van der Waals loop associated with inflated, compact droplets and deflated, string-like droplets~\cite{LSF,mags} respectively are characterised  by the crossover of the mean radius of gyration $R_g(n_c, \Delta p_c, T)= n_c^{\nu} F({\bar p} n_c^{2 \nu})$, where $F(x)$ is a crossover function. The value of $\nu$ is $0.64$ for branched polymers and the $-$ve and $+ve$ ${\bar p}$ branches crossover to $\nu = 0.5$ and $1$ respectively. In Fig.\ref{rgvsnc}{\bf a} we recover this crossover of the droplet shapes for $\phi \leq \phi^*$ for the two corresponding branches of the van der Waals loop as in the unstrained solid. As $\phi$ increases, however, the giant percolating cluster dominates the droplet configuration space and it is impossible to obtain compact or string-like droplets; all droplets now show branched polymer behaviour (Fig.\ref{rgvsnc}{\bf b}) with $R_g \sim n_c^{0.64}$ typical of a system above a critical point. This disappearance of the transition in shapes coincides with the disappearance of the van der Waals loop in ${\bar p} - {\rho_c}$ plane. 

\subsection*{Nucleation of defects}
\begin{figure}[h]
\begin{center}
\includegraphics[width=0.5\textwidth]{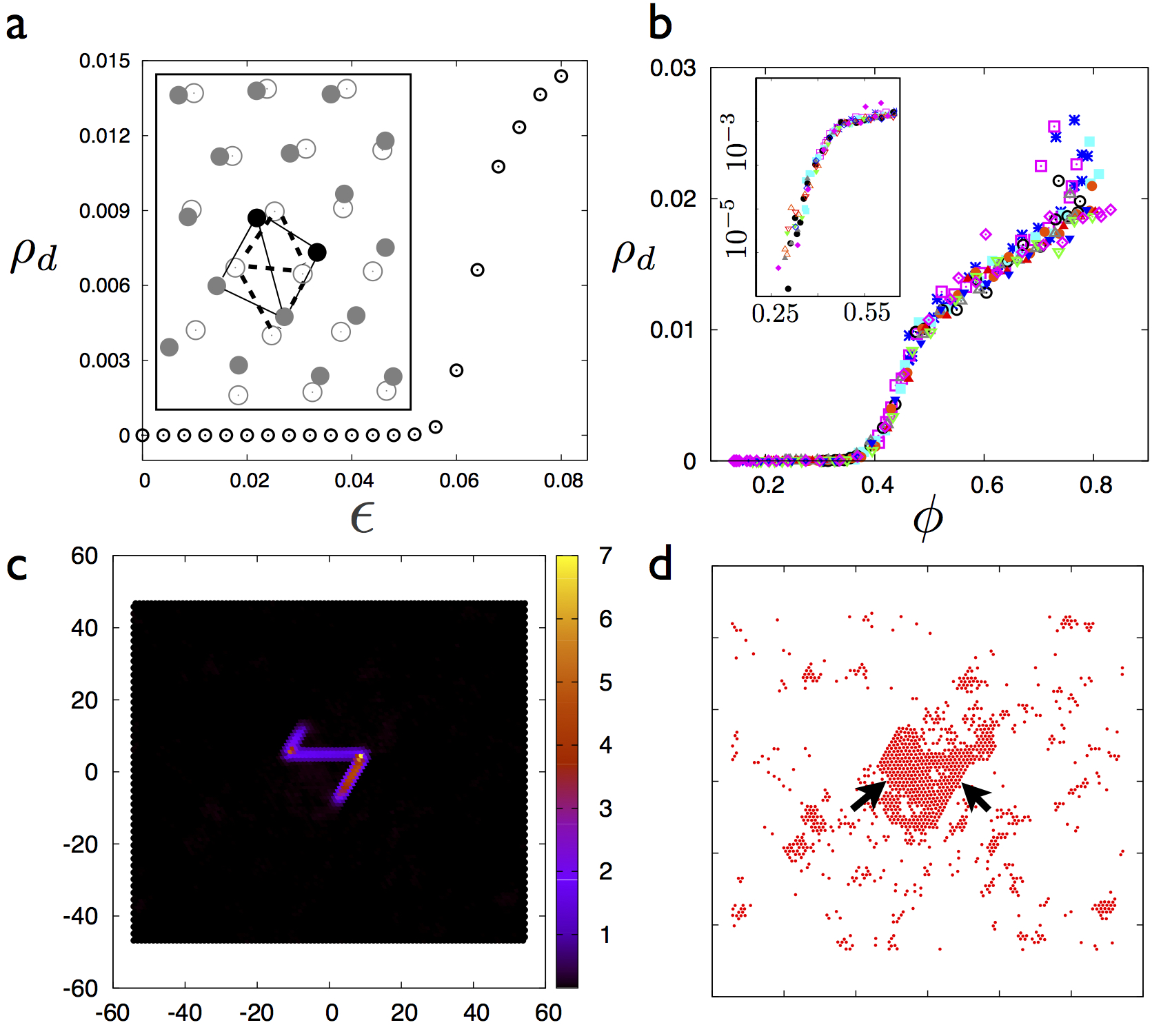}
\end{center}
 \caption{ {\bf a} Plot of the defect concentration $\rho_d$ (particles with $7$ Delaunay neighbours)  vs. $\epsilon$ in the 2D-LJ solid at $\rho = 0.99$ showing sharp  growth of $\rho_d$ for $\epsilon > \epsilon^*$. {\it Inset} shows a particle neighbourhood at two different values of $\epsilon = .039$ (open circles) and $.068$ (filled gray circles), showing the nucleation of a defect pair (black filled circles). Dashed and the bold lines shows change in topology of the local Delaunay neighbourhood with strain. {\bf b} Plot of $\rho_d$ vs. $\phi$ for all $\rho$ and $\epsilon$ showing data collapse onto a single curve with $\rho_d$ increasing sharply at $\phi^*$. {\it Inset} shows the same data in semi-log axes to emphasise the small $\phi$ region. Symbols: same as in Fig.\ref{perc}a. {\bf c} Particles coloured (colour key) according to the value of the local non affine parameter $\chi$ for an isolated dislocation-anti-dislocation pair in a triangular crystal. {\bf d} The same configuration as in {\bf c} showing non-affine particles with $\chi > \chi_{\rm cut}$ (red filled circles). Majority of non-affine particles are associated with the dislocation dipole (black arrows).} 
\label{LTnD}

\end{figure}

Non-affine droplets are associated with coordination number changing deformations. To show this, we track the local concentration of defect pairs by counting the number of nearest neighbours of particles using a local Delaunay triangulation (see Fig.\ref{LTnD}a inset).   In Fig.\ref{LTnD}a, we plot the density of defect pairs $\rho_d$ as a function of the external strain for a LJ solid. While a small number of dislocation pairs (dipoles) always exist within a solid at finite temperatures\cite{dis-number2}, there is a sudden increase in $\rho_d$ as soon as $\epsilon$ crosses a critical value. This increase in defect pair concentration coincides with the critical percolation transition of the non-affine clusters. That this is not a mere numerical coincidence is obvious from the scaling collapse of  $\rho_d$ for all $\rho$ and $\epsilon$  onto a single curve when plotted against the non-affine number fraction $\phi$ in Fig.\ref{LTnD}b. 

We have obtained identical results also at $T = 0.35$ with data collapse occurring for {\em both} $\rho$ and $T$ implying that Fig.\,\ref{LTnD}b represents an {\em universal} relation between $\rho_d$ and $\phi$ for all $\rho$, $T$ and $\epsilon$\, for the 2D-LJ solid. A similar phenomenon occurs for the 3D-LJ solid as well. We discuss these results later in this work. 

Dislocations are preferentially nucleated within the percolating cluster. Also, each defect pair is dressed with an extended region of non-affine particles contributing to the system-spanning, non-affine cluster. We show this by introducing a dislocation pair deleting a row of $20$ atoms in a triangular crystal of $100\times100$ LJ atoms at $\rho = 0.97$. After relaxation to obtain a reference configuration, $\chi$ is calculated during a MD run  at $T = 0.4$ (Fig. \ref{LTnD}c and d). Percolation of non-affine clusters and nucleation of dislocation pairs occur hand-in-hand at the percolation point. This is the main result of the present work.

\subsection*{Elastic nonlinearity, heterogeneity, anelasticity}

How does this percolation transition affect mechanical properties of the solid ? 
%
Increased non-affine fluctuations at defect sites is expected to reduce the local elastic modulus\,\cite{zhou}. The average elastic modulus therefore involves $\rho_d$, the defect concentration and therefore decreases significantly from the linear Hooke's law behaviour as soon as $\rho_d$ becomes large \cite{hashin}. This results in three observable features in the mechanical response of the solid. 

\begin{figure}[h]
\begin{center}
\includegraphics[width=0.49\textwidth]{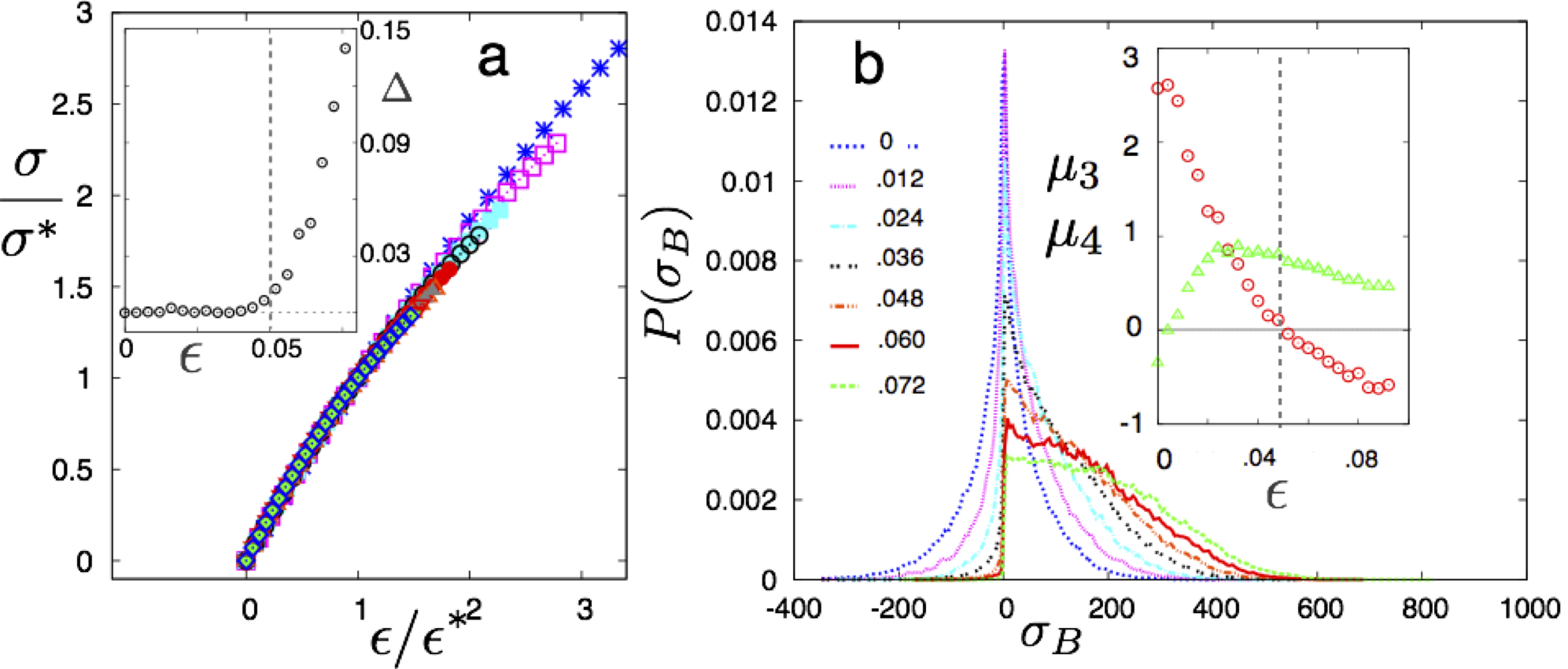}
\end{center}
 \caption{
 {\bf a} Scaled stress $\sigma/\sigma^*$ vs. $\epsilon/\epsilon^*$ in the 2D-LJ solid at $T=0.4$. The symbols, which have the same meaning as in Fig.~\ref{perc}a, shows values for different $\rho$. {\it Inset} shows non-linearity $\Delta$ vs. $\epsilon$. The vertical dashed line marks $\epsilon^*$. {\bf b} Distribution of local stress, $P(\sigma_B)$ obtained from $40\times40$ sub-blocks. At small strains the distribution is Lorentzian. As $\epsilon$ crosses $\epsilon^*$, however, spatially correlated dislocation structures develop making $P(\sigma_B)$ assymetric and broad. {\it Inset} shows variation of skewness (triangles) $\mu_3 = \langle (\sigma - \bar \sigma)^3 \rangle /\Sigma^3$ and kurtosis (circles) $\mu_4 = \langle (\sigma - \bar \sigma)^3 \rangle /\Sigma^4 - 3$ of $P(\sigma_B)$ with $\epsilon$. Note that $\mu_4$ changes sign close to $\epsilon = \epsilon^*$; $\mu_3 = \mu_4 = 0$ for a Gaussian. Here $\bar \sigma $ and $\Sigma$ are the mean and standard deviation of $P(\sigma_B)$.}
\label{stress}

\end{figure}

The first concerns the appearance of a slight nonlinearity in the stress-strain relation 
of the bulk solid. In Fig.\ref{stress}a we show quasi static  stress-strain curves for the LJ solid at $T=0.4$ for $0.9 < \rho < 0.99$ scaled by $\sigma^*$ and $\epsilon^*$ the stress and strain respectively at percolation. Notice that for all $\epsilon \le \epsilon^*$, the curves collapse  to the trivial, linear, response while for larger strains the response becomes non-linear. The non-linearity $\Delta$ defined as the deviation of the stress from a linear fit to the small $\epsilon$ behaviour becomes significant  at precisely $\epsilon^*$ where the non-affine droplets percolate. Similar behaviour, described later, is observed for the 3D-LJ solid beyond the percolation point.

\begin{figure}[h!]
\centerline{\includegraphics[width=0.52\textwidth]{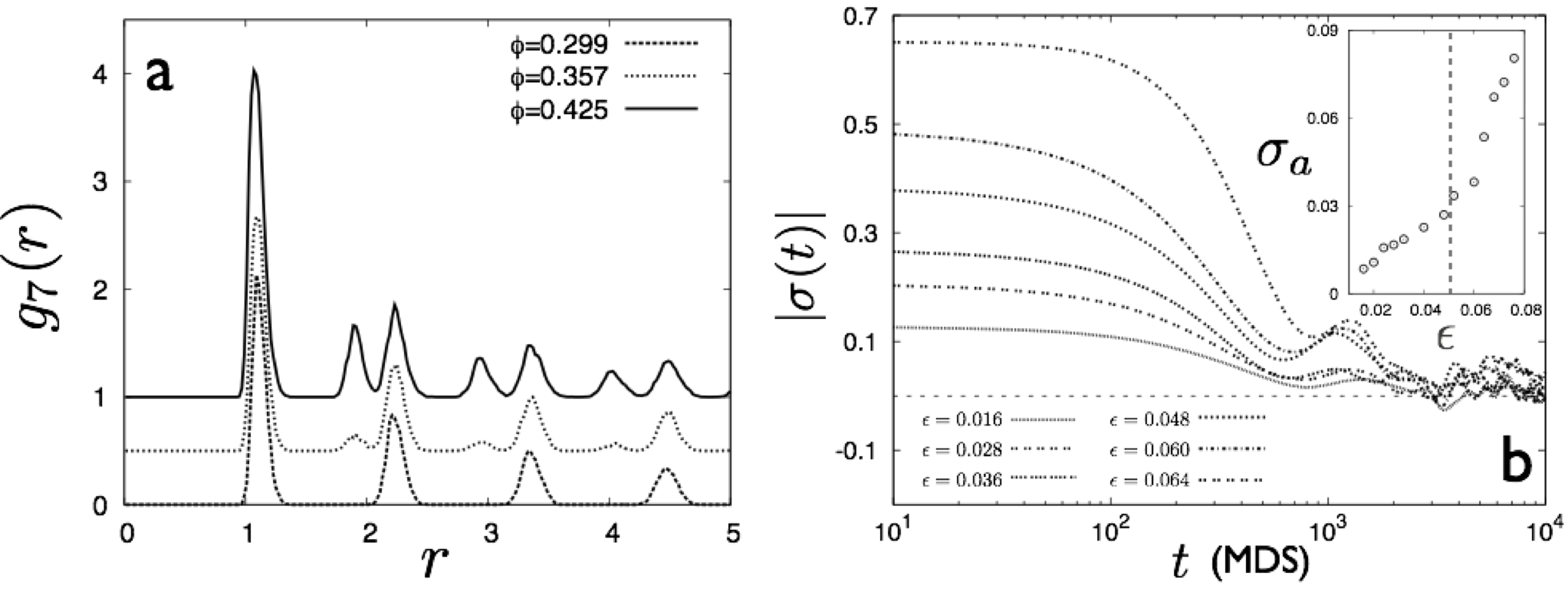}}
\caption{
{\bf a.} The pair distribution function of $7-$ coordinated particles for a $\rho = 0.99$ 2DLJ solid at $T = 0.4$ showing the increase of spatial correlation of the defects as the solid is strained across the percolation transition at $\phi \approx 0.4$. {\bf b.} Anelastic response of the solid at  $\rho=0.99$. A few strained configurations ($\epsilon = 0.016,0.028,0.036,0.048,0.060,0.064$)  are quenched suddenly to zero strain by rescaling the atomic coordinates to study the relaxation of stress $\sigma(t)$ over time $t$. The stress responds by first decreasing rapidly to a negative value whose magnitude increases with the initial $\epsilon$ and then slowly relaxing to zero from below. For the minimally strained solid, the relaxation is essentially exponential. If the initial strain is around $\epsilon^*$,  the dynamical response shows a second relaxation which is complex and non-monotonic. The inset shows the  dependence of the average stress $\sigma_a$ during the relaxation process on the value of the  initial $\epsilon$ before $\sigma(t)$ decreases to $0$. While $\sigma_a$ grows linearly for small $\epsilon$, its growth becomes rapid as $\epsilon$ crosses $\epsilon^*$
}
\label{rho-d-gr}
\end{figure}

The other mechanical feature that manifests at the percolation transition, is the onset of {\em elastic heterogeneity} in the solid. As the number of defects, characterised by pairs of $5-7$ coordinated particles in the 2D-LJ solid, increases dramatically at percolation with defects getting nucleated preferentially within the percolating cluster, one expects that defects also get spatially correlated. Correlated defects cause the local elastic response to be different from the bulk. This may be seen from a local elastic analysis, Fig.\ref{stress}b. Here we plot the local stress distribution $P(\sigma_B)$ by dividing our simulation cell into rectangular sub-blocks \cite{BlockAna} and obtaining local stresses $\sigma_B$ by averaging the virial over the block. While the stress distribution of a perfect crystal is expected to be a Gaussian, that arising from a small concentration of spatially uncorrelated ideal dislocation dipoles is known to be a Lorentzian, $P(\sigma) = \pi^{-1}\sigma_0/(\sigma^2 + \sigma_0^2)$ with $\sigma_0 = D \rho_{\rm dip}/2 \pi$ where $D$ is a constant proportional to the strength of the dipoles, $\rho_{\rm dip}$ is the dipole density\cite{defects}. Our solid at $T = 0.4$ does contain a small defect density $\rho_d \sim 10^{-4}$ even at $\epsilon = 0$. As $\epsilon$ increases, so does $\rho_d$ making $P(\sigma_B)$ broader. At $\epsilon > \epsilon^*$, however, there is a dramatic increase of $\rho_d$ and large numbers of correlated defect pairs are produced. The transformation of $P(\sigma_B)$ can be quantified by calculating the $3^{rd}$ and $4^{th}$ order moments which show that at the percolation point the stress distribution changes from being a sharp Lorentzian to a flat, ``sub-Gaussian'' form. Position correlation between defects can also be seen directly from the pair distribution function of $7-$coordinated particles for solids. We have plotted this quantity for three values of $\phi$ across the transition in FIg.\ref{rho-d-gr}. The increase of correlations is obvious.  



Lastly, crystalline solids, as well as bulk metallic glasses, often show complex, time-dependent, stress relaxation behaviour at values of external strains much below the yield point without permanent plastic deformation. To study the an-elastic response of the LJ solid, we quench a few configurations of our LJ solid, which were equilibrated at several values of $\epsilon$, by rescaling the particle coordinates and the boundary to zero strain. Once initial transients die out, the solid enters a regime of slow relaxation. The nature of this relaxation of the stress, $\sigma(t)$,  which is driven by the motion and subsequent annihilation of defect pairs of opposite sign, depends on the number of such defects and therefore on the initial strain. If the initial strain (and consequently, the defect density) is small, this regime is not very prominent. On the other hand, if the initial $\epsilon$ is large, there are a large number of defects which reorganise over a much longer timescale. In this case, therefore, the second relaxation, as shown in Fig.\ref{rho-d-gr}b is complex and shows non-monotonicity. We characterise this relaxation by plotting the time average of the stress $\sigma_a = \tau^{-1}\int_0^\tau\,dt\, \sigma(t)$ for various values of the initial strain and observe a sharp change of slope at $\epsilon^*$. Since defect pairs are formed at the percolation critical point, the origin of non-monotonic relaxation is traced to the percolation transition at $\phi^*$. The anelastic regime, grows with lowering $\rho$ (and increasing $T$) as expected.  

\subsection*{Effect of temperature}
The percolation threshold depends on $\phi$ the fraction of non-affine particles and not separately on temperature, density and strain. The main thesis of our work is that mechanical properties, such as the defect nucleation threshold, is determined by percolation of non-affine droplets. Therefore, mechanical properties of the solid should depend on external parameters such as temperature, density and strain only through the dependence of $\phi$ on these parameters. We have shown this in Fig.~\ref{perc}a and b for the combination of density and strain at a fixed temperature. It is clear that $\phi$ decreases with increasing density and increases with strain. The percolation threshold at fixed temperature therefore moves to higher strain with increasing density and lower density with increasing strain. If, on the other hand, the temperature is lowered at fixed density one should go to higher strains to achieve percolation. Similarly, the percolation threshold, for fixed strain, should moves to lower densities at lower temperatures.
\begin{figure}[h]
\centerline{\includegraphics[width=0.4\textwidth]{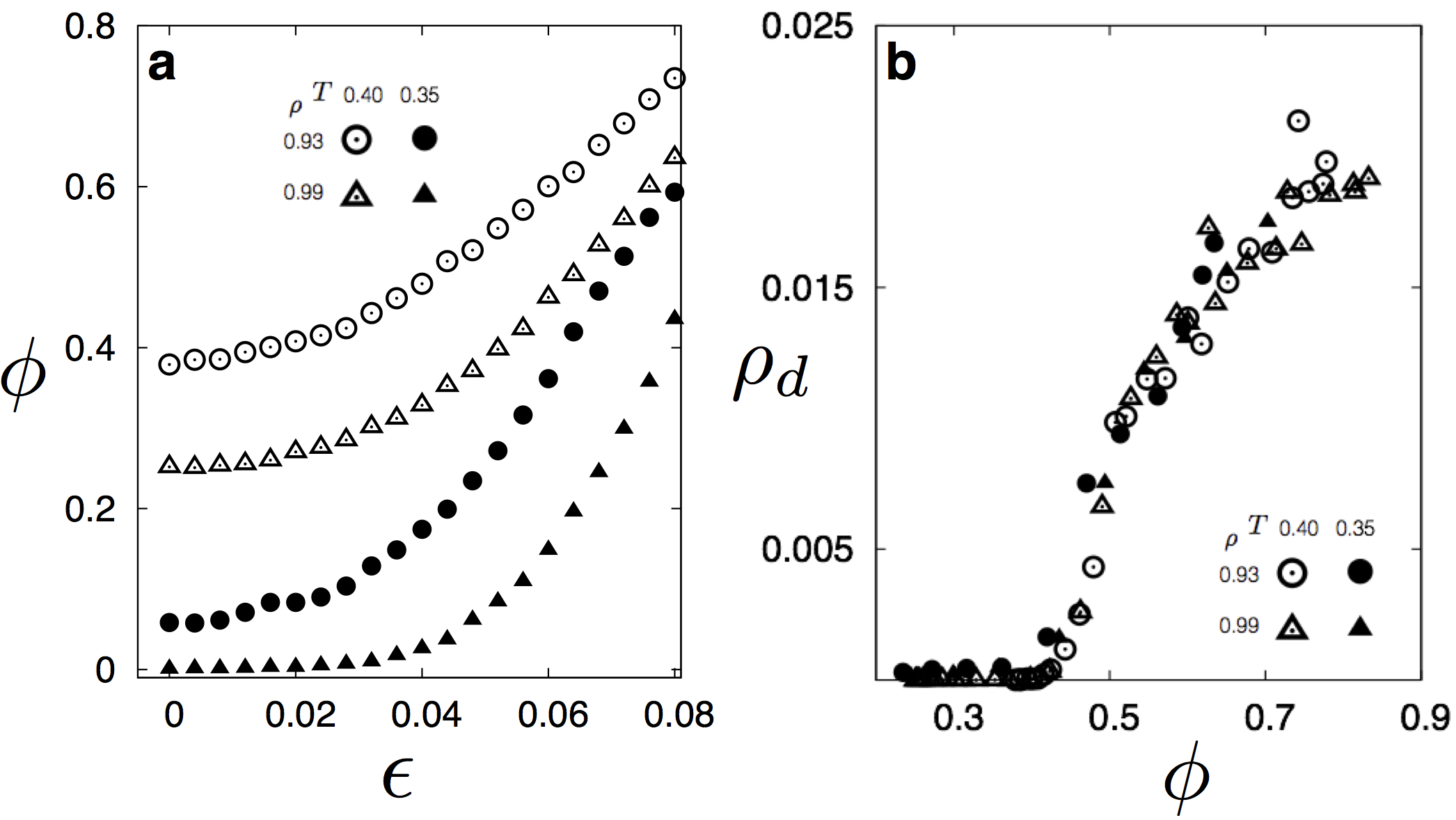}}
\caption{
{\bf a.}\,Plot of the non-affine fraction $\phi(T,\rho)$ for two temperatures $T = 0.35\, \&\, 0.4$ and at $\rho = 0.93\, \& \,0.99$ for the 2DLJ solid. {\bf b.}\,Data collapse plot of $\rho_d$, the defect density, plotted against the non-affine fraction $\phi$.  Note that the data for $T=0.4$ has also been plotted in Fig. 3 b of the manuscript. }
\label{collapse}
\end{figure}
To put to test the validity of our results at other temperatures, we have computed $\rho_d$ for a 2DLJ solid with $\rho = 0.93$ and $0.99$ at both $T=0.35$ and $T=0.40$. We show that $\phi(T,\rho)$ in Fig.~\ref{collapse}a behaves exactly according to our expectations. We also show in Fig.~\ref{collapse}b that one can collapse the data for $\rho_d$ as the solid is quasi-statically strained at different temperatures and densities using the non-affine fraction $\phi$.  The relation shown in Fig.~\ref{LTnD}b is therefore, indeed, an {\em universal}  curve, valid for all densities, strains and temperatures. We show below that the 3D-LJ solid behaves similarly.

\subsection*{Extension to three dimensions}
\begin{figure*}[ht]
\centerline{\includegraphics[width=.7\textwidth]{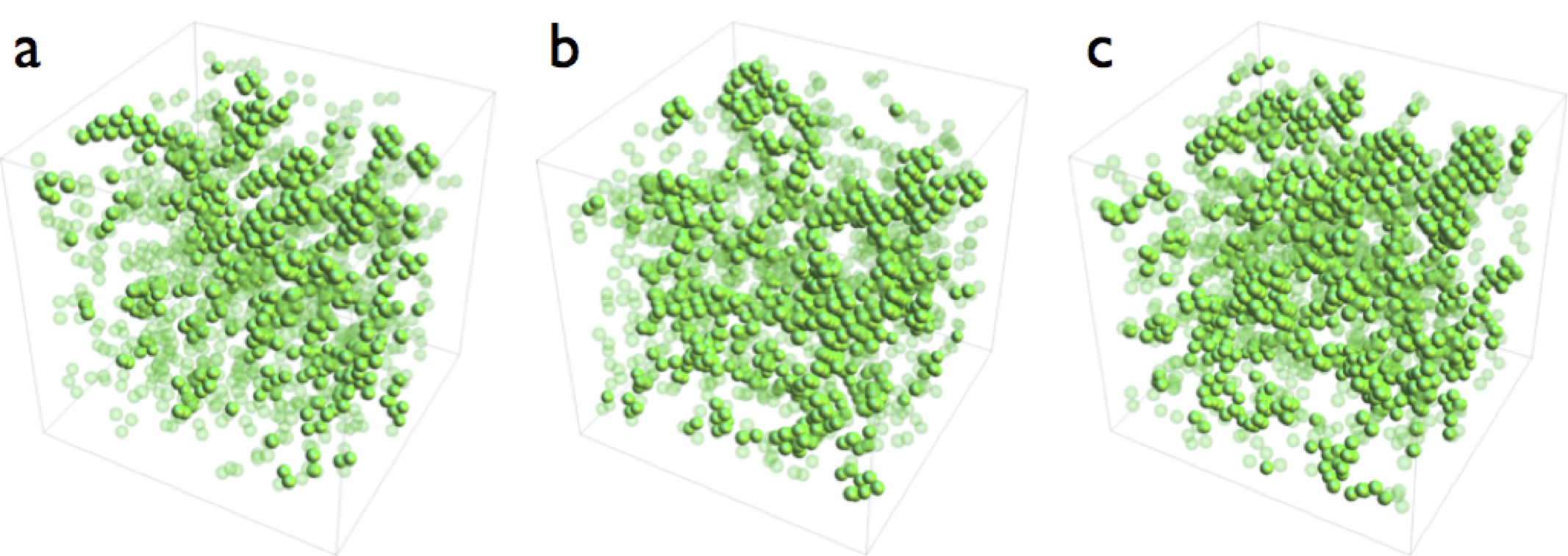}}
\caption{
 {\bf a}-{\bf c} Non affine particles in 3D-LJ, FCC solid of $16384$ particles at $T = 0.7$, $\rho = 1.5$ for $\epsilon =  .02$, $\epsilon = .04$ and $\epsilon = .06$ respectively. Particles with less than $2$ non-affine neighbours have been made translucent for clarity. Percolation occurs for $\epsilon \gtrsim .04$ at this density where $\phi \simeq 0.19$. 
}
\label{3d}
\end{figure*}

\begin{figure}[h]
\centerline{\includegraphics[width=.5\textwidth]{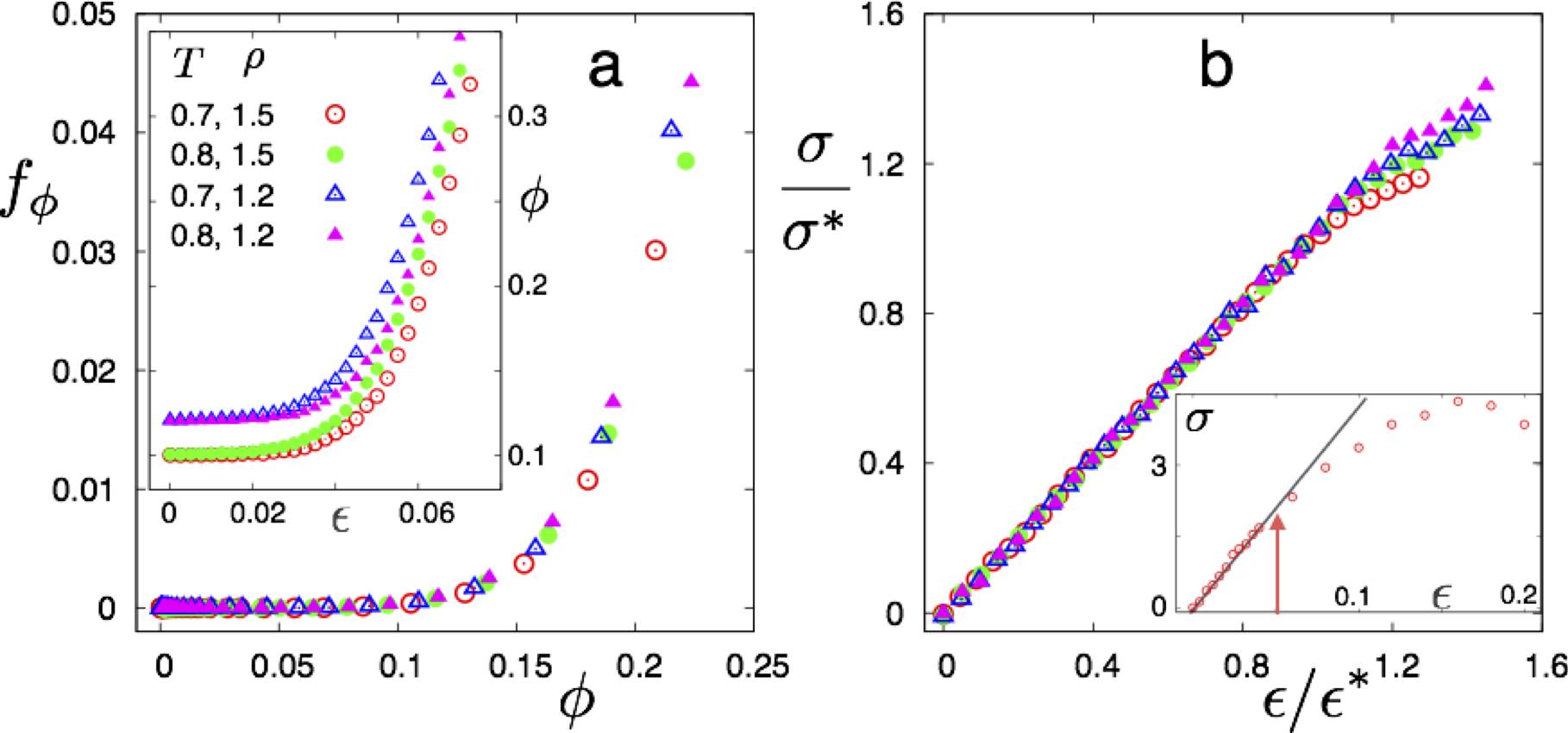}}
\caption{{\bf a} The percolation transition for non affine particles in the 3D-LJ, FCC solid at two values of $T = 0.7 \& 0.8$ and $\rho = 1.5 \& 1.2$. {\it Inset} shows the corresponding $\phi(\rho, T)$ curves.  {\bf b} Scaled stress-strain relations for the sheared solids are shown in {\bf a}; symbols have the same meaning as in {\bf a}. Non-affine particles percolate the solid at $\epsilon/\epsilon^* = 1$. {\it Inset} shows the $\sigma$ vs $\epsilon$ curve for $T=0.7$ and $\rho = 1.5$. The arrow marks $\epsilon^*$ and the solid line marks the linear Hooke's law behaviour.}
\label{3d-2}
\end{figure}

We have shown that in the 2D-LJ solid at any density and temperature, defect nucleation happens at a particular value of strain $\epsilon^*(\rho,T)$ where non-affine clusters percolate. The mechanical response of the solid $\sigma(\epsilon)$ is trivial below $\epsilon^*$ and the stress-strain curves collapse onto each other when scaled by $\epsilon^*$ and $\sigma^*$. The cause for nontrivial mechanical response is the nucleation of defects which increase rapidly for $\epsilon > \epsilon^*$ where clusters containing non-affine particles percolate. Are our results valid only for the 2D-LJ solid or does this have more general validity? To check this, we have looked at the 3D-LJ solid which has a close packed FCC structure. Since fluctuations generally decrease with increase in the number of dimensions, a na\"ive argument may suggest that non-affine fluctuations may not have much of a role in determining the mechanical response of the 3D-LJ solid. 
 The situation is however more subtle,  since in higher dimensions {\it there are more modes that are non-affine}. For example, choosing a neighbourhood which includes only nearest neighbours, the triangular lattice with $6$ neighbours, features $4$ affine (volume, uniaxial extension, shear and rotation) and $8$ non-affine modes \cite{saswati}, while in three dimensions, the FCC lattice with $12$ neighbours has $9$ affine and $27$ non-affine modes. It is possible that the larger number of available non-affine modes more than compensate for the reduction of fluctuations due to the increased dimensionality. 
To check this we simulate the 3D-LJ solid (see {\bf Methods}) at LJ reduced temperatures of $T = 0.7$ and $0.8$ and at $\rho = 1.5$ and $1.2$. The fraction of non-affine particles $\phi$ is obtained using a procedure analogous to what we use in 2D.  The solids are subjected to deviatoric strain $\epsilon = \epsilon_{xx} - \epsilon_{yy} - \epsilon_{zz}$ and the conjugate stress $\sigma$, $\phi$ and $f_\phi$ obtained as in 2D. As expected, $\phi$ increases with $\epsilon$ for fixed $\rho$ and $T$ and decreases at fixed $\epsilon$ as either $\rho$ is decreased or $T$ increased similar to the 2D case; $f_\phi$ vs. $\phi$ is an universal curve with a percolation transition at $\phi \approx 0.19$, close to the expected value for 3D site percolation~\cite{percolation} (see Fig.~\ref{3d-2}a). Using these results, we obtain $\epsilon^*$ and $\sigma^*$, the critical strain and conjugate stress at percolation and use this to scale the stress-strain curve shown in Fig.~\ref{3d-2}b. Again, as in 2D, the stress-strain response becomes non-trivial only beyond the point that non-affine droplets percolate. 
%
%
%
%

Before we end this section, we point out a remarkable aspect of the data shown in Fig.\ref{3d-2}. In 3D, identification of the dominant topological defect is non-trivial. In the FCC lattice, lattice dislocations, frequently decompose into partials producing stacking faults, making a computation of $\rho_d$ from lattice configurations much more difficult that the simple counting of near neighbour coordination which we did in 2D. The nucleation of stacking faults governed by their energy determines the dominant deformation mechanism in FCC crystals. Therefore  considerable effort is needed to define and compute stacking fault energies~\cite{fcc-plast}. Our results may offer an alternative. We have shown that the non-affine fraction $\phi$ provides an equally good measure for significant departure of the lattice from the topology of the ideal reference and may be used to understand pre-yielding phenomena in higher dimensional solids too. Of course, a computation of $\phi$ and $f_\phi$ needs data for instantaneous particle positions which may be available only in computer simulations or from colloidal solids.

\section*{Discussion}
In summary, we have unearthed a hidden mechanical critical point associated with the percolation of non-affine droplets which is intimately tied to the onset of complex mechanical response in a crystalline solid. This  finding is quite unexpected in a system as familiar as a crystalline solid. This transition does not  manifest in mechanical or thermodynamic properties in the bulk, but subtly reveals itself in at least four ways : 

(i) Dislocations, nucleated upon shear, are associated with a spatially extended {\em non-affine cloud}. The spatial scale is determined by two coarse-graining parameters $\chi_{\rm cut}$ and $\Lambda$ (see {\bf Methods}). It is this cloud of non-affinity that eventually percolates across the sample, the percolation transition being fairly robust to small changes in $\chi_{\rm cut}$ and $\Lambda$. 

(ii) The mechanical critical point is not apparent in bulk thermodynamics  but in the restricted (finite size) thermodynamics of the localized droplets. 
The droplet fluctuations represent regions 
where the solid explores nearby minima in the free energy corresponding to metastable glassy or liquid configurations \cite{yukalov}. 
At an algorithmic level, since the droplets are small and transient, we require a special algorithm to distinguish them from the overwhelmingly large contribution of normal fluctuations in the equilibrium solid.

(iii) This mechanical critical point is not reflected in the bulk elastic properties of the solid, but shows up in 
the distribution of local elastic stresses and local elastic constants and the emergence of 
shows strong spatial heterogeneity invisible in bulk elastic experiments.

(iv) Finally, the mechanical critical point is linked to the onset of a strong nonlinear dynamical response of the crystalline solid. Hysteretic response of the stress during strain cycling, slow creep relaxation of the strain under constant load and relaxation of stress {\em below} yield point are all manifestations of so-called ``an-elastic'' behaviour \cite{hysteresis} caused by non-linearities in the elastic response, elastic heterogeneities and time-dependent, slow, defect reorganization \cite{anelasticity}. 
These features are precisely what characterizes the hidden mechanical critical point. 
  
The fact that the emergence of plastic behaviour in {\it both} crystalline and amorphous solids is associated with the percolation of localised non-affine deformations\cite{schall}, suggests that there might be a common language which describes the mechanical response of solids in general. Indeed a similar percolation transition for quasi-statically strained amorphous solids in the athermal ($T=0$) limit has been recently described in \cite{itamar}. 

The hidden critical point betrays itself in our MD simulations only through local properties of non-affine droplets identified using carefully chosen cut-offs and thresholds. Before we end, we speculate on whether the critical point may be revealed by changing system parameters or by introducing novel forces so that it begins to affect bulk thermodynamics~\cite{frenkel}. We have shown in Ref.~\cite{saswati} that an external field $h_\chi$ which couples to $\chi$ may be tuned to control the equilibrium value of $\chi$ and hence the defect density. This field may be introduced in MD simulations as well as realised experimentally in a colloidal solid using holographic optical tweezer techniques\cite{tweeze}. The results presented here for $h_\chi = 0$ may be a reflection of an equilibrium phase transition in the full $h_\chi - \sigma$ space occurring for $h_\chi \ne 0$. In the future, we would also like to study how such hidden critical points influence dynamical behaviour such as avalanches and intermittency~\cite{karmakar}.

{\small 
\section*{Methods}
\subsection*{Simulation details}
\subsubsection*{Two dimensions:}~
We prepare a triangular arrangement of $22,500$ identical particles interacting pairwise via the LJ potential.
Equilibration of our 2D model system is carried out at the fixed $T$ using the dissipative particle dynamics (DPD) protocol implemented within the LAMMPS package (http://lammps.sandia.gov) and preserving both volume and number. Solids at each $\rho$ and $T$ are equilibrated with $10^6$ molecular dynamics time-steps (MDS) where each time-step equals $10^{-4}$ LJ time units. Temperature fluctuations measured at equilibrium is of the order of $1$ in $10^{3}$. Pure shear is then applied to the system following the protocol: $L_x^\prime = (1+e)L_x$ and $L_y^\prime = (1-e)L_y$; $L_x$ and $L_y$  denoting the box lengths in $X$- and $Y$-directions respectively with primed quantities representing the same after shear, $e$ is the shear step set to $0.002$. After each shear step, the system is equilibrated for $4\times10^4$ MDS. Configurations for the analysis is stored in the last $1\times10^4$ MDS in regular intervals of $20$ MDS. This cumulative shearing process is continued till the solid fails at the yield point $\epsilon_c$. The strain rate associated with the process therefore equals to $10^{-3}$ per LJ unit. The sound velocity propagating through the solid $c_s=\sqrt{G/\rho}$ where $G$ is the elastic modulus for pure shear and $\rho$ is the density of the medium. For $\rho=0.99$, ($L_x=174.96$ and $L_y=151.51$ in LJ units), sound at velocity $c_s \sim 6.12$ in LJ units takes approximately $27$ LJ time unit to travel across the simulation box in $X$-direction. Considering this, we refer to our shear protocol as {\em quasi-static}. The model solid at various densities is strained using the same quasi-static shear protocol. 
%
\subsubsection*{Three dimensions:}~The number of particles is $16384$ i.e. $16\times16\times16$ cubic unit cells of the FCC lattice, each containing $4$ basis atoms. The same LJ interactions and the same ensemble with fixed number, volume and $T$ (using the DPD thermostat) are used. The solid is in pure shear with $L_x$ being extended by $\epsilon$, and $L_y$ and $L_z$ shortened such that the volume (and density) is preserved. The strain step is $0.002$ as before. At every strain step the solid is equilibrated for $1.5\times10^5$ MD steps with a time step of $1.\times10^{-3}$ in LJ units. Data is collected for the next $5.\times10^4$ at every $500$ steps.

\subsection*{Identification of droplets}
We identify non-affine droplets using the a prescription described in \cite{pre}. Briefly, at a fixed state point (fixed $T$, $\rho$ and $\epsilon$), we calculate $\chi$ for each particle within a neighbourhood $\Omega$, defined by $r_{ij} \leq \Lambda$ for all configurations using the $T=0$ defect free lattice at the same $\rho$ and $\epsilon$ as reference. We next obtain the distribution $P(\chi)$. We identify a fixed cutoff $\chi_{cut}$ above which the fluctuations are deemed anharmonic. All particles in a single snapshot having $\chi > \chi_{cut}$ are tagged and tagged particles residing within the first nearest neighbor shell of another tagged particle belong to the same cluster; any cluster having at least $7$ particles is identified as a non-affine droplet. Note that $\Lambda$ and $\chi_{cut}$ are parameters which are chosen such that $\Lambda = 2.5$ (2D) or $1.1$ (3D) and $\chi_{cut}$ excludes $90\%$ of the weight of the first peak in $P(\chi)$. Small deviations of these parameters do not change our results substantially.  

\subsection*{Local thermodynamic quantities}
We describe below our procedure for obtaining the local thermodynamic quantities for the droplets.
\subsubsection*{Local density $\rho_c$:}~Once a droplet has been identified, we include one extra layer of particles so that the Voronoi nearest neighbors of all the particles within the cluster are included (whether non-affine or not). Then we count the number of Delaunay triangles using {\em GEOMPACK}\,(code available at \verb+http://people.sc.fsu.edu/~jburkardt/f_src+ \verb+/geompack1+). The correct density of the droplet is then given by half of the ratio between number of particles $n_c$ in the droplet and the total area of all the triangles $\rho_c\equiv n_c/(2A_c)$. 
\subsubsection*{Local pressure $p_c$:}~This is computed using the virial $\langle{\bf F}_{ij} \cdot {\bf r}_{ij}\rangle$ where ${\bf F}_{ij}$ and ${\bf r}_{ij}$ are the nearest neighbor forces and distances respectively for particles $i$ and $j$ belonging to the droplet. The average $\langle...\rangle$ is over all $n_c$. The droplets are characterized by a distribution of $\rho_c$  and excess pressures, $\Delta p_c \equiv p_{c}-p$, where $p$ is the mean pressure of the surrounding solid. 
\subsubsection*{Block stress $\sigma_B$:}~ To compute $\sigma_B$, we divide the simulation cell into sub-blocks $B$ of equal size. The stress is computed by averaging the virial for all particles $N_B$ in the blocks, over all the blocks and finally, over many independent configurations. The components of the stress distribution are calculated as $(\sigma_B)_{\alpha \gamma}=(1/2 N_B)\sum_{k,l\in~B}(x_\alpha^{(l)}-x_\alpha^{(k)})f_\gamma^{(kl)}$ where $k,l$ are particle indices while $\alpha,\gamma$ denotes the cartesian components of the stress. 
}

\noindent{\bf Acknowledgments}
We thank F. Spaepen, P. Sollich and S. Karmakar for discussions. TD acknowledges support from the Collective Interactions Unit of the Okinawa Institute of Science and Technology Graduate University. SG thanks CSIR (India) for support from a Senior Research Fellowship. 

\noindent{\bf Author contributions}
MR and SS conceived of the project and wrote the paper. TD did the necessary calculations, designed the specific algorithms and helped in writing the paper. SG checked the calculations and helped in designing algorithms and obtaining the data.  

\noindent{\bf Additional information}
\noindent Supplementary information accompanies this paper at http://www.nature.com/scientificreports

\noindent Competing financial interests: The authors declare no competing financial interests.


\onecolumn
\newpage
\appendix
\section{
Supplementary Information for:}
{\Large \textbf{Pre-yield non-affine fluctuations and a hidden critical point in strained crystals}}
\vspace{0.6cm}

\noindent \textbf{Tamoghna Das,\textit{$^{a,b}$} Saswati Ganguly,\textit{$^{b}$} Surajit Sengupta\textit{$^{\ast}$}\textit{$^{c}$} and Madan Rao\textit{$^{d\ddag}$}}\vspace{0.5cm}

\noindent{\textit{$^{a}$~Collective Interactions Unit, OIST Graduate University,1919-1 Tancha, Onna-son, Okinawa, Japan - 904-0495}}

\noindent{\textit{$^{b}$~Centre for Advanced Materials, Indian Association for the Cultivation of Science, Jadavpur, Kolkata 700032, India }}

\noindent{\textit{$^{c}$~TIFR Centre for Interdisciplinary Sciences, 21 Brundavan Colony, Narsingi, Hyderabad 500075, India.}\,email:\,surajit@tifrh.res.in}

\noindent{\textit{$^{d}$~Raman Research Institute, C.V. Raman Avenue, Bangalore 560080, India }}

\subsection{The non-affine parameter $\chi$, identification of non-affine clusters and Probability $P(\chi, \epsilon)$ of non-affine displacements as a function of strain}

In this section, we give details of how we  identify non-affine droplets in the 2D-LJ and 3D-LJ solids under pure shear \cite{this}. Our procedure follows closely that used in Ref.\cite{that}. Since all the particles in the ideal solid are identical, non-affine fluctuations in crystals arise solely from thermal effects. Non-affine regions therefore are composed of those particles which suffer somewhat large random deviations from their ideal lattice positions due to thermal motion. We identify particles in non-affine regions by computing the instantaneous non-affine parameter $\chi$.

To obtain the non-affine parameter $\chi$ for a particle $i$ we use a metric neighbourhood $\Omega$ around particle $i$ defined using a cutoff size $\Lambda = 2.5$ corresponding to the range of interactions and minimise the error $\chi = \sum_j [{\bf r}_i - {\bf  r}_j - (1 + \epsilon)({\bf R}_i - {\bf R}_j)]^2$ with respect to choices of the strain tensor $\epsilon$. Here ${\bf r}_i$ and ${\bf R}_i$ correspond to the instantaneous and reference positions of $i$, respectively and the sum is over the neighbours of $i$.  A similar procedure was used in Ref.\cite{falk} to obtain a measure of non-affine displacements in amorphous solids. However, we modify the definition used in \cite{falk} in two important ways. 

Firstly, we compute  $\chi$ using a fixed reference $\{{\bf R_i}\}$. For the case where pure shear strain is zero (as in Ref.\cite{that}) we have used the $T=0$ ideal crystal lattice as reference. Where the solid is under pure shear, we have used as reference the ideal triangular lattice deformed by the uniform pure shear $\epsilon$ for our reference configuration. Secondly, the number of neighbours in $\Omega$ between the instantaneous and reference neighbourhoods is allowed to be different, accounting for {\em inclusions} within $\Omega$ which changes the metric coordination number. If there are no inclusions, the computed $\chi$ and $\epsilon$ are identical in meaning to that in \cite{falk}. If, however, there are inclusions, then extra contributions to both $\epsilon$ and $\chi$ arise from a change in coordination. In this case, the fitted $\epsilon$ looses its meaning as a local strain and $\chi$ computes non-affineness in both displacements and the number of inclusions. For the parameters that we study, the number of inclusions is always at most one.   

In Fig.\ref{pchi} we plot the probability distribution $P(\chi)$ obtained for a solid of $\rho = 0.99$ for strains (pure shear) $0.0 \leq \epsilon \leq 0.072$. For a solid with fixed neighbors, $P(\chi)$ is an unimodal distribution\cite{sas}. When the number of neighbours is allowed to vary between the reference and the deformed configurations then $P(\chi)$ has additional peaks corresponding to each inclusion. As $\epsilon$ increases, the weight of $P(\chi)$ shifts to the peaks at large $\chi$. From the distribution $P(\chi)$ we determine a cut-off $\chi_{\rm cut}$ above which particle displacements are defined to be {\em non-affine}.  This is defined as in Ref.\cite{that} by obtaining $P(\chi)$ for an equivalent harmonic system and equating $\chi_{\rm cut}$ with the value above which $P(\chi)$ for the harmonic lattice has negligible weight. This effectively discards $90\%$ of the weight of the first peak in $P(\chi)$ (Fig.\ref{pchi}{\bf a}) as coming from trivial harmonic fluctuations. All particles in a single snapshot having $\chi > \chi_{cut}$ are tagged and tagged particles residing within the first nearest neighbour shell of another tagged particle belong to the same cluster; any cluster having at least $7$ particles is identified as a non-affine droplet. 

\begin{figure}[h]
\centerline{\includegraphics[width=0.6\textwidth]{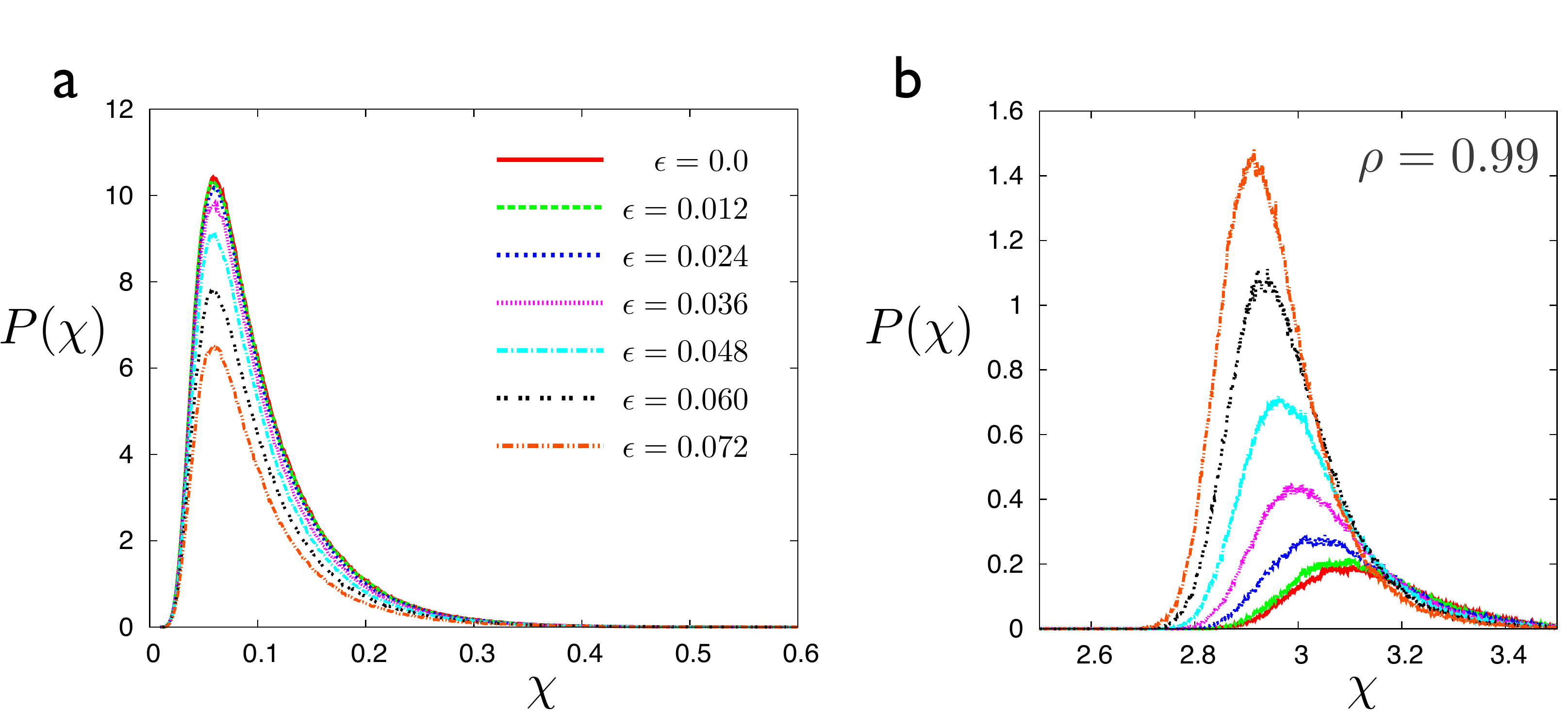}}
\caption{
Normalized distribution of non-affine parameter $P(\chi)$ is shown for the case of $\rho=0.99$ at $T=0.4$. The two different figures {\bf a} \& {\bf b}, shows two different ranges of the same distribution. Note the difference in the scale of the $y$-axis. $P(\chi)$ is essentially bimodal for the unstrained case as in Ref.\cite{that}. Applied strain $\epsilon$ enhances non-affine displacements as observed by the increase in height of the second peak ($2.5\leq\chi\leq4.5$) at the expense of decreasing first peak ($\chi\leq0.5$).  Close observation to the local configurations reveals the fluctuations corresponding to the first peak keeps the numbers of neighbors intact within the cut off radius $\Lambda = 2.5$, the interaction range, the subsequent peaks arise when there are  inclusions within the reference volume $\Lambda$ changing the number of neighbors.
}
\label{pchi}
\end{figure}
Our results are robust with respect to small variations of the cut-offs $\Lambda$ and $\chi_{\rm cut}$. Increasing $\Lambda$ increases the non-affine fraction while increasing $\chi_{\rm cut}$ decreases it.  Large deviations of these numbers therefore shifts the focus from the fluctuations of interest namely all non-harmonic, non-affine fluctuations occurring within a neighbourhood of the dimensions of the interaction volume of the system. 


A similar situation exists in 3D. We present the normalised distribution of $\chi$ as a 3D-LJ solid (face centred cubic) is strained via pure shear protocol (Fig.\ref{3d} (a) \& (b)) as in 2D.  In this case, we use a {\em metric} neighbourhood with $\Lambda = 1.1$ to obtain $\chi$ similar to the 2D-LJ case. As the system under consideration is deep in the solid region ($\rho=1.5, T = 0.7$), the distribution $P(\chi)$ for the unstrained case ($\epsilon = 0.0$) is unimodal and breaks up into a bimodal distribution once sheared. First peak of the distribution decreases upon increasing shear at cost of higher, co-ordination number changing, non-affineness emerging in the system which contribute to the second peak. The situation is qualitatively in fair accord with its 2D counterpart. These distributions have been used to set the cut-off $\chi$ using the same criterion as in 2D and thus to define the non-affine clusters presented in the main text for the 3D case.

\begin{figure}[h!]
\centerline{\includegraphics[width=0.6\textwidth]{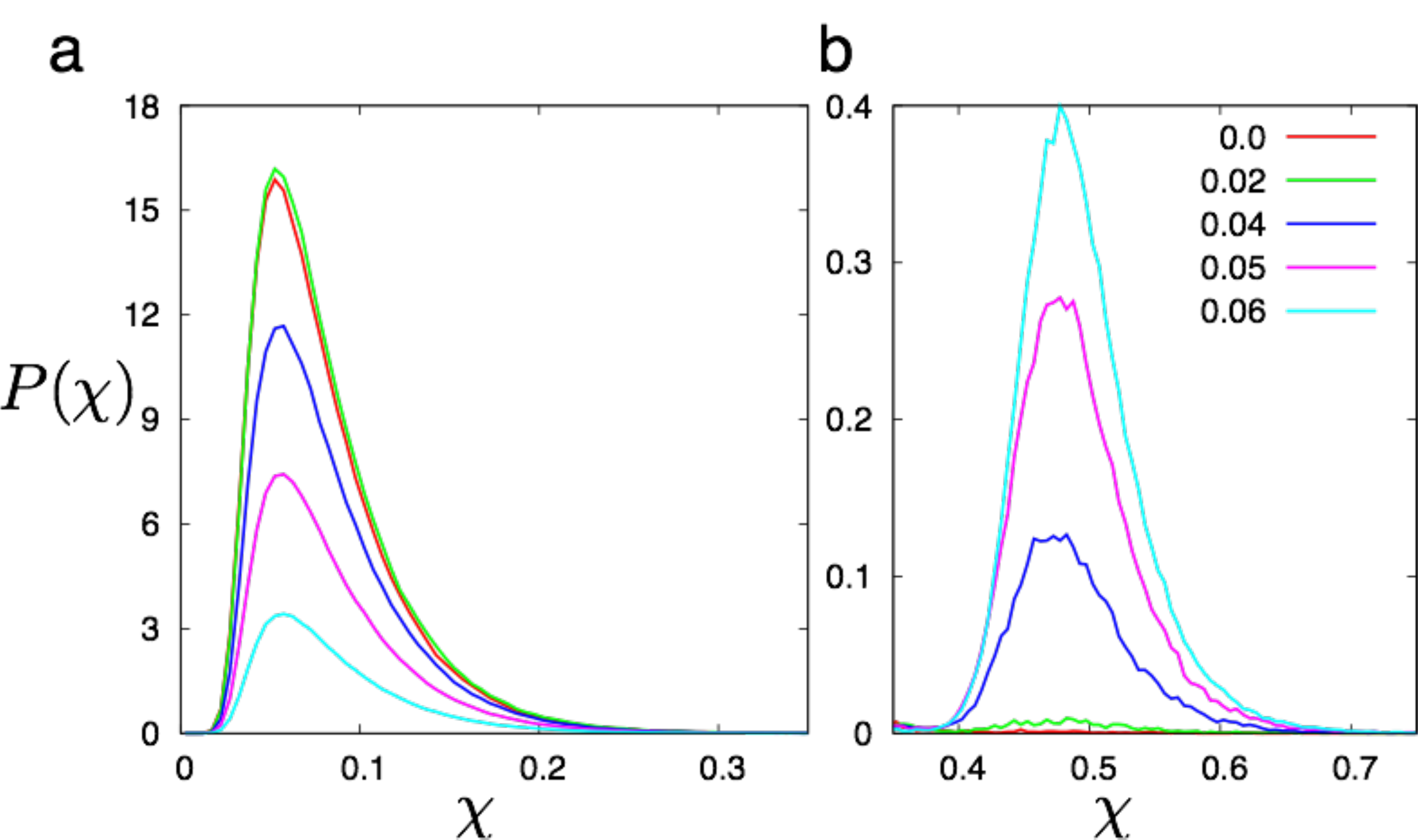}}
\caption{
Increase in non-affineness upon shearing a 3D LJ solid (FCC) at $\rho = 1.5$ and $T = 0.7$ is shown in {\bf a} \& {\bf b}. Both of them are part of the same distribution, but shown with separate axes for clarity. }
\label{3d}
\end{figure}

\subsection{Local thermodynamics of non-affine droplets in other model two dimensional solids}

In the present paper \cite{this} and our earlier work \cite{that} we have related non-affine displacement fluctuations in the Lennard Jones (LJ) solid to droplet fluctuations from ``nearby'' metastable states such as the liquid and glass. How general is this association and what are the limits of its validity ? 
To test the {\it generality} of this result, we  carry out our analysis on another model solid where the particles interact with a different potential (for which it is known that both liquid and glassy free energy minima exist). To test the {\it  limits} of this result, we study a model solid where, by construction, there are no locally stable phases except for the crystalline solid. 
\begin{figure}[h]
\centerline{\includegraphics[width=0.6\textwidth]{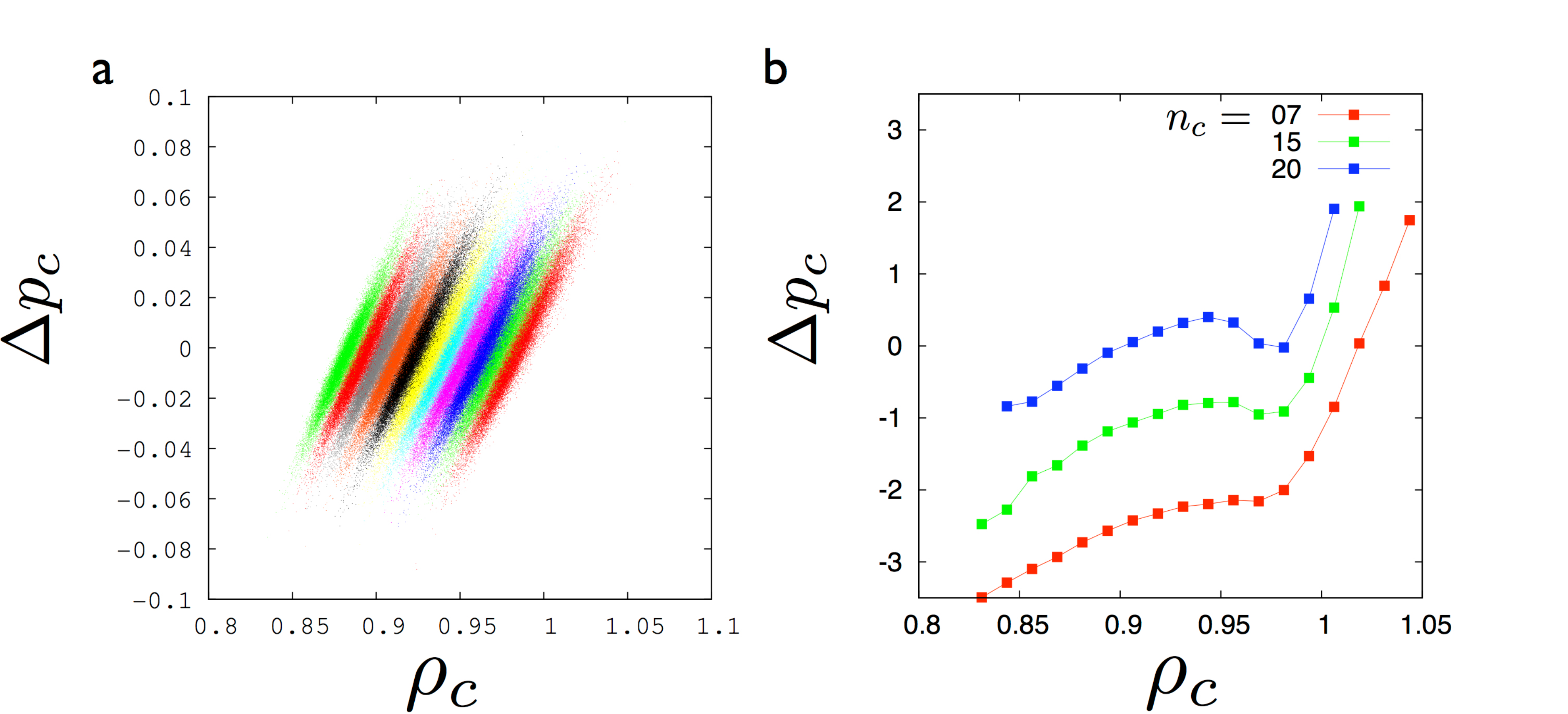}}
\caption{
{\bf a} A scatter plot showing the local pressure difference $\Delta p_c$ as a function of the local density $\rho_c$ for ``non-affine'' droplets in the harmonic triangular net obtained using a lower cut-off $\chi_{\rm cut} = 0.2$. The different colours correspond to solids at different densities obtained by applying a hydrostatic compression (or expansion) on the $T=0$ solid with unit lattice parameter. The results show that in this system non-trivial droplet fluctuations do not exist and all fluctuations are normally distributed around the mean density and pressure. The slope of the plots correspond to the bulk modulus determined by the spring constant $K$ which is the same for all the solids.  
{\bf b} Plot of local thermodynamic quantities $\Delta p_c$ vs. $\rho_c$ for non-affine droplets in the WCA system at a $T^{*} = 0.4$ for droplets of size $n_c = 7, 15$ and $20$. The prominent van der Waals loop in this quantity is similar to the corresponding plots (see Fig. 2 in \cite{this} or Fig. 7(b) in \cite{that}) for the LJ system.
}
\label{harmonic}
\end{figure}

\subsubsection{Harmonic triangular net} 
The harmonic triangular net where particles in a two dimensional triangular lattice are connected to their nearest neighbors by harmonic springs (spring constant, $K$) is the simplest model of a {\it network} solid. The potential energy for this system is given by,
\begin{equation}
V = \sum_{\langle ij \rangle} K (\vert {\bf r}_i  - {\bf r}_j \vert - r_0)^2.
\end{equation}
Here, ${\bf r}_i$ is the position vector of the $i-$th particle, $r_0$ is the equilibrium bond length and the sum extends over the nearest neighbor shell.  Since defects, which change the connectivity of the lattice are not allowed by construction, such a solid never melts and the free energy surface in configuration space is particularly simple having just one quadratic minimum corresponding to the $T=0$ ideal triangular lattice. 

We have obtained the probability distribution of the non-affine parameter $P(\chi)$ in this system using the same procedure as used in the LJ system and using a neighborhood which includes the first two neighbor shells. The control parameter in our model is the lattice parameter $l_p = \sqrt{2/\sqrt{3} \rho}$ where $\rho$ is the density of the solid and we hold the temperature $T$, the equilibrium bond length $r_0$ and $K$ fixed at unity. The reference configuration is the ideal triangular lattice with the same lattice parameter. As in the LJ solid, the $P(\chi)$ is bimodal and resembles Fig.\ref{pchi} and Fig. 2 in \cite{that} for the unstrained solid at values of $K$ low enough such that particles enter or leave the reference volume $\Omega$. When non-affine droplets are defined using the same cutoff $\chi_{cut}$ as in \cite{that}, the local thermodynamics of these droplets as given by the $\Delta p_c$ vs. $\rho_c$ scatter plots are entirely unremarkable and are normally distributed around the average density of the solid. This is shown in Fig.\ref{harmonic}\,{\bf a}. Further, this behaviour is, in this system, independent of the cut-off $\chi_{\rm cut}$ used. We do not see any evidence of  a van der Waals loop in this case. 

This is  an important null result which shows that systems where there are no liquid like or glassy minima, do not have nontrivial local thermodynamics of the non-affine droplets and these droplets therefore do not correspond to fluctuations of any metastable minima. 

\subsubsection{The Weeks-Chandler-Anderson solid}
In contrast to the harmonic solid, the system of particles interacting with the purely repulsive Weeks, Chandler and Anderson (WCA) potential \cite{wca} does undergo melting \cite{wca-melt} and glass
\cite{wca-glass} transitions. This potential is closely related to the LJ potential and is given by, 
\begin{eqnarray}
V(r) & = & 4 (r^{-12} - r^{-6}) + 1\,\,\,\,\,\,r \leq 2^{1/6} \\ \nonumber
       & = & 0\,\,\,\,\,\,\,\,\,\,\,\,\,\,\,\,\,\,\,\,\,\,\,\,\,\,\,{\rm otherwise}
\end{eqnarray}
Unlike the LJ system, the WCA potential does not support a gas phase and there is no liquid - gas critical point. 

We simulate a $N=1024$ system of WCA particles in the NVT ensemble for several densities and a few temperatures. The configurations obtained are treated exactly in the same way as in the LJ solid and non-affine droplets are identified from the plots of $P(\chi)$. Local thermodynamics of these droplets then yield  $\Delta p_c$ vs. $\rho_c$ curves which are qualitatively similar to the LJ case in \cite{that}. 

We have thus shown that there is strong correlation between the appearance of non-trivial local thermodynamics of the non-affine droplets and the presence of metastable liquid or glass like phases. While not a complete proof, taken together, our results strongly suggest  that non-affine fluctuations in a crystalline solid are droplet fluctuations from nearby metastable minima.



\begin{thebibliography}{}
\bibitem{dieter} Dieter G., {\it Mechanical Metallurgy}, (McGraw-Hill, New York, 1961)
\bibitem{haasen} Cahn R. W.  \& Haasen P., {\it Physical Metallurgy} 4$^th$ Ed. (North-Holland, Amsterdam, 1996)
\bibitem{fcc-plast} Jo, M. et. al. Theory for plasticity of face-centered cubic metals., Proc. Natl. Acad. Sc. {\bf 111}, 6560-6565 (2014).
\bibitem{anelasticity} Nowick, A. S. \& Berry, B. S. {\it Anelastic relaxation in crystalline solids}, (Academic Press, New York, 1972)
\bibitem{ana-glass1}Luborsky, F. E. eds. {\it Amorphous Metallic Alloys} (Butterworths, London, 1983)
\bibitem{ana-glass2}Tomida, T. \& Egami, T. Molecular-dynamics study of structural anisotropy and anelasticity in metallic glasses, \prb {\bf 48}, 3048-3057 (1993).
\bibitem{Hirth}Hirth, J. P. \& Lothe, J. {\it Theory of Dislocations}, (Kreiger, Malabar, 1982)
\bibitem{ane2}Nowick, A. S. \& Heller, W. R. Dielectric and anelastic relaxation of crystals containing point defects., Adv. in Phys. {\bf 14}, 101-166 (1965)

%
%
%
\bibitem{Argon} Argon, A. Plastic deformation in metallic glasses., Acta Met. {\bf 27}, 47-58 (1979).
\bibitem{spaepen} Spaepen, F. A microscopic mechanism for steady state inhomogeneous flow in metallic glasses., Acta Met. {\bf 25}, 407-415 (1977).
\bibitem{FL}Falk, M. L. \& Langer, J. S.  Dynamics of viscoplastic deformation in amorphous solids., \pre, {\bf 57}, 7192-7205 (1998). 
%
%


%
%
%
\bibitem{saswati} Ganguly, S. Sengupta, S. Sollich, S. P. \& Rao, M. Nonaffine displacements in crystalline solids in the harmonic limit., \pre {\bf 87}, 042801(2013).
\bibitem{pre} Das, T. Sengupta S. \& Rao, M. Nonaffine heterogeneities and droplet fluctuations in an equilibrium crystalline solid., \pre {\bf 82}, 041115 (2010). 
\bibitem{ums} Frenkel, D. \& Smit B. {\em Understanding Molecular Simulations} (Academic Press, San Deigo 2002).
\bibitem{LJ} Barker, J. A. Henderson, D. \& Abraham, F. F. Phase diagram of the two-dimensional Lennard-Jones system; Evidence for first-order transitions., Physica {\bf 106A}, 226-238 (1981).
\bibitem{LSF} Leibler, S. Singh, R. R. P.  \& Fisher, M. E. Thermodynamic behavior of two-dimensional vesicles., \prl  {\bf 59}, 1989-1992 (1987).
\bibitem{mags} Maggs, A. C. Leibler, S. Fisher, M. E. \& Camacho, C. J. Size of an inflated vesicle in 2 dimensions., \pra {\bf 42}, 691-695 (1990). 
\bibitem{brpl} Hsu, H.-P. Nadler, W. \& Grassberger, P. Statistics of lattice animals., Comput. Phys. Commun. {\bf 169}, 114 (2005).
\bibitem{zahn}Zahn, K. Wille, A. Maret, G. Sengupta, S. \& Nielaba, P. Elastic Properties of 2D Colloidal Crystals from Video Microscopy., Phys. Rev. Lett. {\bf 90}, 155506 (2003).
\bibitem{3dlj3pt} Hansen, J.-P. \& Verlet, L. Phase Transitions of the Lennard-Jones System., Phys. Rev. {\bf 184}, 151-161 (1969).
\bibitem{percolation} Stauffer, D. \& Aharony, A. {\it Introduction to Percolation Theory} 2$^{nd}$ Edition, (Taylor and Francis, London, 1994).
\bibitem{binder} Binder, K. Theory of First-Order Phase Transitions., Rep. Prog. Phys. {\bf 50}, 783-859 (1987).
\bibitem{dis-number2} Sengupta, S. Nielaba, P. \& Binder, K. Elastic moduli, dislocation core energy, and melting of hard disks in two dimensions., \pre {\bf 61}, 6294-6301 (2000).
\bibitem{zhou} Zhou, C. et al. Dislocation-induced anomalous softening of solid helium., Phil. Mag. Lett. {\bf 92}, 608-616 (2012).
\bibitem{hashin} Hashin, Z. \& Shtrikman, S. A  variational approach to the theory of the elastic behaviour of multiphase materials., J. Mech. Phys. Sol. {\bf 11}, 127-140 (1963).
\bibitem{BlockAna} Rovere, M. Nielaba, P. \& Binder, K. Simulation studies of gas-liquid transitions in 2 dimensions via a subsystem-block-density distribution analysis,  Z. Phys. {\bf 90}, 215-228 (1993).
\bibitem{defects} Csikor F. F. \& Groma, I. Probability distribution of internal stress in relaxed dislocation systems.,\prb {\bf 70}, 064106 (2004).
\bibitem{yukalov} Yukalov, V. I. Phase-transitions and heterophase fluctuations., Phys. Rep. {\bf 208}, 395-489, (1991).
\bibitem{hysteresis} Laurson L. \& Alava, M. J. Dynamic Hysteresis in Cyclic Deformation of Crystalline Solids., \prl {\bf 109}, 155504 (2012).
\bibitem{schall} Schall, P. Weitz, D. A. \& Spaepen, F. Structural rearrangements that govern flow in colloidal glasses, Science {\bf 318}, 1895-1899 (2007).
\bibitem{itamar} Chikkadi, V. et al., Percolating Plastic Failure as a Mechanism for Shear Softening in Amorphous Solids.,  arXiv:1312.4136.
\bibitem{frenkel} ten Wolde P. R.  \& Frenkel, D. Enhancement of protein crystal nucleation by critical density fluctuations., Science,  {\bf 277}, 1975-1978 (1997).
\bibitem{tweeze} Irvine, W. T. M. Hollingsworth, A. D. Grier D. G. \& Chaikin, P. M. Dislocation reactions, grain boundaries, and irreversibility in two-dimensional lattices using topological tweezers. Proc. Nat. Acad. Sc., {\bf 110}, 15544-15548 (2013).
\bibitem{karmakar} Karmakar, S. Lemaitre, A. Lerner, E. and Procaccia, I. Predicting Plastic Flow Events in Athermal Shear-Strained Amorphous Solids., \prl, {\bf 104}, 215502 (2010).
\end{thebibliography}

\begin{thebibliography}{}

\bibitem{this} Accompanying manuscript T. Das, S. Ganguly, S. Sengupta and M. Rao.


\bibitem{that}T. Das, S. Sengupta and M. Rao, Phys. Rev. E {\bf 82}, 041115 (2010).

\bibitem{falk} M. L. Falk and J. S. Langer, Phys. Rev. E {\bf 57}, 7192 (1998).

\bibitem{sas}S. Ganguly, S. Sengupta, P. Sollich, M. Rao, Phys. Rev. E {\bf 87}, 042801 (2013).

\bibitem{wca}J. D. Weeks, D. Chandler, and H. C. Andersen, J. Chem. Phys. {\bf 54}, 5237 (1971).

\bibitem{wca-melt}A. Ahmed and R. J. Sadus, Phys. Rev. E. {\bf 80}, 061101 (2009).

\bibitem{wca-glass}T. Kawasaki, T. Araki and H. Tanaka, Phys. Rev. Lett. {\bf 99}, 215701 (2007).

%
%


\end{thebibliography}
\end{document}